\documentclass[english,manuscript,A4,english, APS,  PRD, nofootinbib,superscriptaddress]{revtex4-1}
\usepackage[T1]{fontenc}
\usepackage[latin9]{inputenc}
\usepackage{color}
\usepackage{babel}
\usepackage{amstext}
\usepackage{amssymb}
\usepackage{esint}
\usepackage[unicode=true,pdfusetitle,
 bookmarks=true,bookmarksnumbered=false,bookmarksopen=false,
 breaklinks=false,pdfborder={0 0 0},pdfborderstyle={},backref=false,colorlinks=true]
 {hyperref}
\hypersetup{
 linkcolor=blue,citecolor=cyan}
\begin{document}
\title{Thermodynamic stability in relativistic viscous and spin hydrodynamics}
\author{Xiang Ren}
\affiliation{Department of Modern Physics, University of Science and Technology
of China, Anhui 230026, China}
\author{Chen Yang}
\affiliation{Department of Modern Physics, University of Science and Technology
of China, Anhui 230026, China}
\author{Dong-Lin Wang}
\email{donglinwang@mail.ustc.edu.cn}

\affiliation{Department of Modern Physics, University of Science and Technology
of China, Anhui 230026, China}
\author{Shi Pu}
\email{shipu@ustc.edu.cn}

\affiliation{Department of Modern Physics, University of Science and Technology
of China, Anhui 230026, China}
\affiliation{Southern Center for Nuclear-Science Theory (SCNT), Institute of Modern
Physics, Chinese Academy of Sciences, Huizhou 516000, Guangdong Province,
China}
\begin{abstract}
We have applied thermodynamic stability analysis to derive the stability
and causality conditions for conventional relativistic viscous hydrodynamics
and spin hydrodynamics. We obtain the thermodynamic stability conditions
for second-order relativistic hydrodynamics with shear and bulk viscous
tensors, finding them identical to those derived from linear mode
analysis. We then derive the thermodynamic stability conditions for
minimal causal extended second-order spin hydrodynamics in canonical
form, both with and without viscous tensors. Without viscous tensors,
the constraints from thermodynamic stability exactly match those from
linear mode analysis. In the presence of viscous tensors, the thermodynamic
stability imposes more stringent constraints than those obtained from
linear mode analysis. Our results suggest that conditions derived
from thermodynamic stability analysis can guarantee both causality
and stability in linear mode analysis.
\end{abstract}
\maketitle

\section{Introduction}

In relativistic heavy ion collisions, two nuclei are accelerated to
speeds close to that of light, collide with each other, and generate
a hot and dense matter known as the quark-gluon plasma (QGP) \citep{BRAHMS:2004adc,PHENIX:2004vcz,STAR:2005gfr,ALICE:2008ngc}.
The evolution of the QGP is well described by relativistic hydrodynamics.
Relativistic hydrodynamics serves as a macroscopic effective theory
for relativistic many-body systems in the long-wavelength and low-frequency
limit. The main equations of relativistic hydrodynamics are the conservation
equations for the energy-momentum tensor and other conserved currents
in the gradient expansion, e.g. the Israel-Stewart theory \citep{Israel:1979wp,Israel:1979},
the extended Baier-Romatschke-Son-Starinets-Stephanov theory \citep{Baier:2007ix},
the Denicol-Niemi-Molnar-Rischke theory \citep{Denicol:2012cn}, and
the more recently established Bemfica-Disconzi-Noronha-Kovtun theory
\citep{Bemfica:2017wps,Kovtun:2019hdm,Bemfica:2019knx,Hoult:2020eho,Bemfica:2020zjp}.
For additional studies and developments, we refer the reader to the
recent review papers \citep{Gavassino:2021kpi,Rocha:2023ilf} and
the references therein.

In the early stages of noncentral collisions, the nuclei possess a
huge initial orbital angular momentum, on the order of $10^{7}\hbar$.
This initial orbital angular momentum is transferred to the spin polarization
of quarks and subsequently to the final-state particles through spin-orbital
coupling. This mechanism leads to the spin polarization of $\Lambda$
and $\bar{\Lambda}$ hyperons and the spin alignment of vector mesons
\citep{Liang:2004ph,Liang:2004xn,Gao:2007bc}. The STAR collaboration
has observed both the global and local polarization of $\Lambda$
and $\bar{\Lambda}$ hyperons \citep{STAR:2017ckg,STAR:2019erd},
as well as the spin alignment of $\phi$ and $K^{0,*}$ mesons \citep{STAR:2022fan}.

On the theoretical side, the global polarization can be well described
by various phenomenological models \citep{Becattini:2007sr,Karpenko:2016jyx,Xie:2017upb,Li:2017slc,Sun:2017xhx,Shi:2017wpk,Wei:2018zfb,Xia:2018tes,Vitiuk:2019rfv,Fu:2020oxj,Ryu:2021lnx,Lei:2021mvp,Wu:2022mkr}
through the combination of the modified Cooper-Frye formula \citep{Becattini:2013fla,Fang:2016vpj}
with hydrodynamic simulations under the assumption that the system
is close to global equilibrium. To understand local polarization,
effects beyond global equilibrium, such as shear-induced polarization
\citep{Liu:2021uhn,Liu:2021nyg,Fu:2021pok,Becattini:2021suc,Yi:2021ryh,Yi:2021unq,Yi:2023tgg,Wu:2023tku},
spin Hall effects \citep{Liu:2020dxg,Fu:2022myl,Wu:2022mkr}, weak
magnetic fields induced polarization \citep{Sun:2024isb} and the
corrections due to the interactions between quarks and back ground
fields \citep{Fang:2023bbw}, need to be considered. Although hydrodynamic
simulations can qualitatively describe local polarization as functions
of azimuthal angle, understanding the dependence on centrality and
transverse momentum remains challenging \citep{STAR:2019erd,ALICE:2021pzu,STAR:2023eck}.
Therefore, it is necessary to consider the evolution of spin during
collisions. Recently established spin hydrodynamics, which integrates
the total angular momentum conservation equation with conventional
relativistic hydrodynamic equations, has been developed from various
theoretical frameworks, such as from effective action \citep{Montenegro:2017rbu,Montenegro:2017lvf},
entropy principle \citep{Hattori:2019lfp,Fukushima:2020ucl,Li:2020eon,Gallegos:2021bzp,She:2021lhe,Hongo:2021ona,Wang:2021ngp,Wang:2021wqq,Cao:2022aku,Hu:2022azy,Biswas:2023qsw},
kinetic theory \citep{Florkowski:2017ruc,Florkowski:2017dyn,Florkowski:2018myy,Weickgenannt:2019dks,Bhadury:2020puc,Weickgenannt:2020aaf,Shi:2020htn,Speranza:2020ilk,Bhadury:2020cop,Singh:2020rht,Peng:2021ago,Sheng:2021kfc,Hu:2021pwh,Weickgenannt:2022zxs,Weickgenannt:2022jes,Weickgenannt:2022qvh,Wagner:2024fhf},
holography \citep{Gallegos:2020otk,Garbiso:2020puw}, and quantum
statistics \citep{Becattini:2023ouz}. For recent reviews on this
topic, see Refs. \citep{Hidaka:2022dmn,Shi:2023sxh,Becattini:2024uha}.

As a fundamental requirement, both conventional relativistic hydrodynamics
and spin hydrodynamics must exhibit causality and stability. In pioneering
works \citep{Hiscock:1985zz,Hiscock:1987zz}, linear mode analysis
was implemented to study the causality and stability of hydrodynamic
systems. Through linear mode analysis, the causality and stability
conditions for various types of hydrodynamics are derived \citep{Hiscock:1985zz,Hiscock:1987zz,Koide:2006ef,Denicol:2008ha,Pu:2009fj,Brito:2020nou,Brito:2021iqr,Sarwar:2022yzs,Daher:2022wzf,Xie:2023gbo,Weickgenannt:2023btk,Shokri:2023rpp,deBrito:2023vzv,Fang:2024skm,Daher:2024bah}.
These conditions establish inequalities that constrain the range of
transport coefficients. Recently, it was found that the conventional
causality criterion \citep{Krotscheck1978CausalityC} used in linear
mode analysis is insufficient to guarantee causality. Consequently,
several studies \citep{Heller:2022ejw,Gavassino:2023myj,Heller:2023jtd,Gavassino:2023mad,Wang:2023csj,Hoult:2023clg}
have proposed new causality criteria that also explore the deep connection
between causality and stability \citep{Bemfica:2020zjp,Gavassino:2021owo,Wang:2023csj}.

Very recently, Ref. \citep{Xie:2023gbo} has systematically studied
the causality and stability for the minimal extended second-order
spin hydrodynamics in the linear mode analysis. Later, Ref. \citep{Daher:2024bah}
also investigates the impact of other second-order terms. It was revealed
that the system appears to be unstable at finite wavelengths, even
though it satisfies asymptomatic stability conditions derived for
both large and small wavelengths \citep{Xie:2023gbo}. To address
this issue, it is essential to explore the stability of spin hydrodynamics
through an alternative approach.

In this work, we apply thermodynamic stability analysis \citep{Hiscock:1983zz,Olson:1990rzl,Gavassino:2021cli,Gavassino:2021kjm},
which is grounded in the second law of thermodynamics and the principle
of maximizing total entropy in equilibrium states \citep{Landau:1980mil},
to spin hydrodynamics. We will derive stability conditions from this
thermodynamic stability analysis and compare them with those obtained
through linear mode analysis.

The structure of this paper is organized as follows: In Sec. \ref{sec:Brief-introduction},
we briefly review thermodynamic stability analysis. Next, we apply
this analysis to conventional relativistic viscous hydrodynamics as
a test case in Sec. \ref{sec:MIS}. In Sec. \ref{sec:spin_hydrodynamics},
we analyze the thermodynamic stability conditions for spin hydrodynamics
and compare the results with those obtained from linear mode analysis.
We conclude with a summary in Sec. \ref{sec:Conclusion}.

Throughout this work, we choose the metric $g_{\mu\nu}=\textrm{diag}\{+,-,-,-\}$
and define the projector $\Delta^{\mu\nu}=g^{\mu\nu}-u^{\mu}u^{\nu}$
with $u^{\mu}$ being the fluid velocity. For an arbitrary tensor
$A^{\mu\nu}$, we introduce the notations $A^{(\mu\nu)}=\frac{1}{2}(A^{\mu\nu}+A^{\nu\mu})$,
$A^{[\mu\nu]}=\frac{1}{2}(A^{\mu\nu}-A^{\nu\mu})$, and $A^{<\mu\nu>}\equiv\frac{1}{2}[\Delta^{\mu\alpha}\Delta^{\nu\beta}+\Delta^{\mu\beta}\Delta^{\nu\alpha}]A_{\alpha\beta}-\frac{1}{3}\Delta^{\mu\nu}(\Delta^{\alpha\beta}A_{\alpha\beta})$.

\section{Brief introduction to the thermodynamic stability \label{sec:Brief-introduction}}

In this section, we briefly review the main idea in Ref. \citep{Gavassino:2021kjm}.
Consider an isolated system near thermodynamic equilibrium, consisting
of a fluid connected to a sufficiently large heat-particle bath. According
to the second law of thermodynamics, the entropy of the entire system,
$S$, must not decrease, i.e., the variation of entropy $\Delta S$
follows: 
\begin{equation}
\Delta S=\Delta S_{F}+\Delta S_{B}\geq0,\label{eq:thermal_stability_01}
\end{equation}
where $S_{F,B}$ stand for the entropy for fluid and bath, respectively.
Equation (\ref{eq:thermal_stability_01}) is the original condition
for the thermodynamic stability.

Now, let us consider conserved quantities $\mathcal{Q}^{a}$ and their
thermodynamic conjugates $\alpha^{a}$ in the system, where $a=1,2,...$
label different conserved quantities. For example, if the total number
is conserved, then $\mathcal{Q}$ and $\alpha$ correspond to the
total number and $\mu/T$, respectively, with $\mu$ and $T$ being
the chemical potential and temperature. While, if the total energy
is conserved, $\mathcal{Q}$ and $\alpha$ are total energy and $-1/T$,
respectively. Then, the variation of entropy can be expressed as
\begin{equation}
dS=-\sum_{a}\alpha^{a}d\mathcal{Q}^{a}.
\end{equation}
The $\mathcal{Q}^{a}$ can be divided as the part for fluid $\mathcal{Q}_{F}^{a}$
and the one for the bath $\mathcal{Q}_{B}^{a}$, with the following
relationship:
\begin{equation}
d\mathcal{Q}_{B}^{a}=-d\mathcal{Q}_{F}^{a}.
\end{equation}
Then the variation of total entropy becomes
\begin{equation}
\Delta S=\Delta S_{F}+\sum_{a}\alpha_{B}^{a}\Delta\mathcal{Q}_{F}^{a}\geq0.\label{eq:delta_S}
\end{equation}
If defining 
\begin{equation}
\Psi\equiv S_{F}+\sum_{a}\alpha_{B}^{a}\mathcal{Q}_{F}^{a},\label{eq:Phi}
\end{equation}
then Eq. (\ref{eq:delta_S}) implies that the function $\Psi$ should
be maximized in the equilibrium state.

One can also define the information current $E^{\mu}$ as
\begin{equation}
E^{\mu}\equiv-\delta s_{F}^{\mu}-\sum_{a}\alpha_{F}^{a}\delta J_{F}^{a,\mu},\label{eq:def_E}
\end{equation}
where $s_{F}^{\mu}$ is the entropy current of the fluid and $J_{F}^{a,\mu}$
is the conserved current associated with $\mathcal{Q}_{F}^{a}$, and
the symbol $\delta$ denotes the small perturbations from the thermodynamic
equilibrium state.

Given that the whole system is near thermodynamic equilibrium and
the heat-particle bath is sufficiently large, we can assume that the
chemical potential and temperature in the fluid are equal to those
in the bath, i.e. $\alpha_{F}^{a}$ for the fluid is approximately
equal to $\alpha_{B}^{a}$ in the bath. Under this assumption, $\alpha_{F}^{a}\delta J_{F}^{a,\mu}$
can be simplified to $\delta(\alpha^{a}J_{F}^{a,\mu})$, where we
do not distinguish between $\alpha_{F}^{a}$ in the fluid and $\alpha_{B}^{a}$
in the bath. Consequently, Eq. (\ref{eq:delta_S}) can be further
written as

\begin{equation}
E\equiv\int d\Sigma\;E^{\mu}n_{\mu}\geq0,\label{eq:E_def}
\end{equation}
holds for an arbitrary spacelike three-dimensional surface $\Sigma$
and its timelike and future-directed normal unit vector $n^{\mu}$.
If the thermodynamic equilibrium state is unique, i.e., determined
solely by the thermodynamic variables, then from Eq. (\ref{eq:E_def})
and the definition of $E^{\mu}$ in Eq. (\ref{eq:def_E}), the information
current $E^{\mu}$ must satisfy the following conditions:
\begin{eqnarray}
\mathrm{(i)} & \  & E^{\mu}n_{\mu}\geq0\textrm{ for any \ensuremath{n^{\mu}} with \ensuremath{n_{0}>0,n^{\mu}n_{\mu}=1}},\nonumber \\
\mathrm{(ii)} & \  & E^{\mu}n_{\mu}=0\textrm{ if and only if all perturbations are zero},\nonumber \\
\mathrm{(iii)} & \  & \partial_{\mu}E^{\mu}\leq0.\label{eq:thermal_critera}
\end{eqnarray}

As a remark, the conditions in Eq. (\ref{eq:thermal_critera}) can
be treated as criteria of thermodynamic stability \citep{Gavassino:2021kjm}.
It has also been found that these criteria in Eq. (\ref{eq:thermal_critera})
can guarantee the causality of the system \citep{Gavassino:2021kjm}.
Moreover, when all these conditions are satisfied in one inertial
frame of reference, the thermodynamic stability conditions in Eq.
(\ref{eq:thermal_stability_01}) are assured across all inertial frames
of Refs. \citep{Gavassino:2021owo,Wang:2023csj}. These criteria provide
us with a novel tool for analyzing the stability and causality of
the system.

\section{Thermodynamic stability of the second order viscous hydrodynamics
\label{sec:MIS}}

In this section, we implement the thermodynamic stability criteria
(\ref{eq:thermal_critera}) to the relativistic second order viscous
hydrodynamics in the gradient expansion. It can be considered as an
example to show the connection between the constraints from the thermodynamic
stability and conventional linear mode analysis. For convenience,
we focus on the quantities for fluid and omit all the lower index
$F$ from now on.

The energy momentum conservation equation reads
\begin{equation}
\partial_{\mu}T^{\mu\nu}=0,
\end{equation}
and the energy momentum tensor in the Landau or energy frame is given
by \citep{landau:1987Fluid}

\begin{eqnarray}
T^{\mu\nu} & = & (e+P)u^{\mu}u^{\nu}-Pg^{\mu\nu}+\pi^{\mu\nu}-\Pi\Delta^{\mu\nu},\label{eq:TmunuIS}
\end{eqnarray}
where $e,P,\pi^{\mu\nu},\Pi$ are energy density, pressure, shear
viscous tensor, and bulk pressure, respectively. Note that the net
baryon number density of the QGP produced in relativistic heavy ion
collisions is negligible \citep{Yagi:2005yb}. For simplicity, the
following discussions are limited in the cases where (baryon) currents
vanish.

In order to compare the constraints from thermodynamic stability and
linear mode analysis, we choose the minimal extension of second order
viscous hydrodynamics \citep{Koide:2006ef}. The corresponding entropy
current is given by
\begin{equation}
s^{\mu}=su^{\mu}-Q^{\mu}+\mathcal{O}(\partial^{3}),\label{eq:entropy_current_01}
\end{equation}
where $Q^{\mu}$ stands for the possible corrections from the second
order. Following Refs. \citep{Israel:1979wp,Israel:1979}, we take
\begin{equation}
Q^{\mu}=\frac{1}{2}\frac{u^{\mu}}{T}(\chi_{\Pi}\Pi^{2}+\chi_{\pi}\pi^{\rho\sigma}\pi_{\rho\sigma}),\label{eq:Q_01}
\end{equation}
as an example. Then the entropy principle $\partial_{\mu}s^{\mu}\geq0$
gives the constitutive equations for $\pi^{\mu\nu}$ and $\Pi$ as
below,
\begin{eqnarray}
\tau_{\Pi}(u\cdot\partial)\Pi+\Pi & = & -\zeta\left[\partial_{\mu}u^{\mu}+\frac{1}{2}\chi_{\Pi}T\partial_{\rho}(u^{\rho}/T)\Pi\right],\nonumber \\
\tau_{\pi}\Delta^{\alpha<\mu}\Delta^{\nu>\beta}(u\cdot\partial)\pi_{\alpha\beta}+\pi^{\mu\nu} & = & 2\eta\left[\partial^{<\mu}u^{\nu>}-\frac{1}{2}\chi_{\pi}T\partial_{\rho}(u^{\rho}/T)\pi^{\mu\nu}\right],\label{eq:Pi_pi}
\end{eqnarray}
where 
\begin{equation}
\zeta,\eta>0,\label{eq:zeta_eta}
\end{equation}
and 
\begin{equation}
\tau_{\Pi}=\zeta\chi_{\Pi},\quad\tau_{\pi}=2\eta\chi_{\pi}.
\end{equation}
For the more comprehensive discussions on the second order theories,
we refer to Refs. \citep{Israel:1979wp,Israel:1979}. Later, we will
compare our results from thermodynamic stability with those from linear
mode analysis \citep{Denicol:2008ha,Pu:2009fj}. The terms proportional
to $\chi_{\Pi},\chi_{\pi}$ on the right-hand side of Eq. (\ref{eq:Pi_pi})
do not appear in the constitutive equations in Refs. \citep{Denicol:2008ha,Pu:2009fj},
but these terms will not contribute to the causality and stability
conditions in linear mode analysis.

Next, we choose local rest frame $u^{\mu}=(1,0)$ and assume the fluid
reaches the thermodynamic equilibrium state, in which $\Pi$ and $\pi^{\mu\nu}$
are zero. For the macroscopic variables $\varphi=(e,u^{\mu},\Pi,\pi^{\mu\nu})$,
we consider the perturbations near the thermodynamic equilibrium,
$\delta\varphi$. We can expand the system in the power series of
$\delta\varphi$. By using the following relationship, 
\begin{eqnarray}
u^{\mu}\delta u_{\mu} & = & -\frac{1}{2}\delta u^{\mu}\delta u_{\mu},\nonumber \\
u_{\mu}\delta\pi^{\mu\nu} & = & -\delta u_{\mu}\delta\pi^{\mu\nu},\nonumber \\
\delta\pi_{\mu}^{\mu} & = & 0,
\end{eqnarray}
we find, 
\begin{eqnarray}
\delta u^{i},\delta\pi^{ij} & \sim & \mathcal{O}(\delta),\nonumber \\
\delta u^{0},\delta\pi^{0i} & \sim & \mathcal{O}(\delta^{2}),\nonumber \\
\delta\pi^{00} & \sim & \mathcal{O}(\delta^{3}).
\end{eqnarray}
With the help of Eq. (\ref{eq:entropy_current_01}), the information
current is then given by \citep{Gavassino:2021cli,Gavassino:2021kjm}

\begin{eqnarray}
E^{\mu} & = & -\delta s^{\mu}+\frac{u_{\nu}}{T}\delta T^{\mu\nu}\nonumber \\
 & = & u^{\mu}\left(\frac{\delta e}{T}-\delta s\right)+\frac{1}{T}\delta u^{\mu}\delta P+\frac{1}{2}\frac{u^{\mu}}{T}(\chi_{\pi}\delta\pi^{\alpha\beta}\delta\pi_{\alpha\beta}+\chi_{\Pi}\delta\Pi\delta\Pi)\nonumber \\
 &  & -\frac{1}{2T}(e+P)u^{\mu}\delta u_{\nu}\delta u^{\nu}-\frac{1}{T}\delta u_{\nu}\delta\pi^{\mu\nu}+\frac{1}{T}\delta u^{\mu}\delta\Pi+\mathcal{O}(\delta^{3}).\label{eq:Emu_01}
\end{eqnarray}
By using the thermodynamic relations,
\begin{equation}
ds=\frac{1}{T}de,\;dP=sdT,\label{eq:thermalRIS}
\end{equation}
we find that
\begin{eqnarray}
\delta s & = & \frac{1}{T}\delta e+\frac{1}{2}\frac{\partial^{2}s}{\partial e^{2}}(\delta e)^{2}+\mathcal{O}(\delta^{3})\nonumber \\
 & = & \frac{1}{T}\delta e-\frac{1}{2T}\frac{\delta P}{e+P}\delta e+\mathcal{O}(\delta^{3})\nonumber \\
 & = & \frac{1}{T}\delta e-\frac{1}{2T}\frac{c_{s}^{2}}{e+P}(\delta e)^{2}+\mathcal{O}(\delta^{3}),\label{eq:delta_s_01}
\end{eqnarray}
where $c_{s}^{2}$ is the speed of sound. Then, $E^{\mu}$ can be
further simplified,

\begin{eqnarray}
E^{\mu} & = & \frac{1}{2T}\frac{u^{\mu}c_{s}^{2}}{e+P}(\delta e)^{2}+\frac{c_{s}^{2}}{T}\delta u^{\mu}\delta e-\frac{1}{2T}(e+P)u^{\mu}\delta u_{\nu}\delta u^{\nu}\nonumber \\
 &  & -\frac{1}{T}\delta u_{\nu}\delta\pi^{\mu\nu}+\frac{1}{T}\delta u^{\mu}\delta\Pi+\frac{1}{2T}u^{\mu}(\chi_{\pi}\delta\pi^{\alpha\beta}\delta\pi_{\alpha\beta}+\chi_{\Pi}\delta\Pi\delta\Pi).
\end{eqnarray}

Let us now impose the three conditions (\ref{eq:thermal_critera})
on $E^{\mu}$. From the definition (\ref{eq:def_E}), we have $\partial_{\mu}E^{\mu}=-\partial_{\mu}\delta s^{\mu}$,
so that the condition (iii) in Eq. (\ref{eq:thermal_critera}) leads
to the inequality (\ref{eq:zeta_eta}), which is consistent with the
requirement from the conventional entropy principle. To analyze the
constraints from the conditions (i) and (ii) in Eq. (\ref{eq:thermal_critera}),
we introduce an arbitrary timelike future-directed vector $n^{\mu}$
with $\ensuremath{n_{0}>0,n^{\mu}n_{\mu}=1}$. After some tedious
and straightforward calculations, we obtain, 
\begin{eqnarray}
\frac{2n_{0}TE^{\mu}n_{\mu}}{e+P} & = & \frac{n_{0}^{2}\tau_{\pi}}{\eta(e+P)}\sum_{i<j}\left[\delta\pi^{ij}-\frac{1}{n_{0}\chi_{\pi}}n_{(i}\delta u_{j)}\right]^{2}\nonumber \\
 &  & +\frac{n_{0}^{2}\tau_{\pi}}{\eta(e+P)}\left[\delta\pi^{11}+\frac{1}{2}\delta\pi^{22}+\frac{1}{2n_{0}\chi_{\pi}}(n_{3}\delta u_{3}-n_{1}\delta u_{1})\right]^{2}\nonumber \\
 &  & +\frac{3n_{0}^{2}\tau_{\pi}}{4\eta(e+P)}\left[\delta\pi^{22}+\frac{1}{3n_{0}\chi_{\pi}}(n_{3}\delta u_{3}+n_{1}\delta u_{1}-2n_{2}\delta u_{2})\right]^{2}\nonumber \\
 &  & +\sum_{i=1}^{5}a_{i}(\delta A_{i})^{2},\label{eq:En_01}
\end{eqnarray}
where the exact expressions for $a_{i}$ and $\delta A_{i}$ can be
found in Appendix \ref{sec:Expressions_aiAi}. Imposing the conditions
(i) and (ii) in Eq. (\ref{eq:thermal_critera}) leads to\footnote{In this work, we assume $e+P>0$, while Ref. \citep{Almaalol:2022pjc}
also explores cases where $e+P<0$. Additionally, we note that the
treatment of $\delta\pi^{\mu\nu}$ in Eq. (\ref{eq:En_01}) is different
with Eq. (C14) of Ref. \citep{Almaalol:2022pjc}, since the number
of independent components of $\delta\pi^{\mu\nu}$ is $5$.}
\begin{eqnarray}
c_{s}^{2},\tau_{\pi},\tau_{\Pi} & > & 0,\nonumber \\
1-c_{s}^{2}-\frac{4\eta}{3\tau_{\pi}(e+P)}-\frac{\zeta}{\tau_{\Pi}(e+P)} & > & 0,\label{eq:MISC2}
\end{eqnarray}
which are exactly the same as those derived from linear mode analysis
in the previous literature \citep{Denicol:2008ha,Pu:2009fj,Pu:2011vr}.

In general, if the baryon or other conserved current is considered,
e.g. $j^{\mu}=nu^{\mu}+\nu^{\mu}$ with $n$ and $\nu^{\mu}$ being
number density and diffusive current, the independent fields become
$\varphi=(e,u^{\mu},\Pi,\pi^{\mu\nu},n,\nu^{\mu})$ \citep{Gavassino:2023qnw}.
In these cases, the thermodynamic relations (\ref{eq:thermalRIS}),
constitutive relations (\ref{eq:entropy_current_01})-(\ref{eq:Pi_pi}),
and information current (\ref{eq:Emu_01}) will be modified. More
constraints for thermodynamic stability would occur and the final
constraints become different with Eq. (\ref{eq:MISC2}). For the general
analysis including baryon currents, one can refer to Refs. \citep{Gavassino:2021cli,Gavassino:2021kjm,Almaalol:2022pjc,Gavassino:2023qnw}.

\section{Thermodynamic stability of spin hydrodynamics \label{sec:spin_hydrodynamics}}

In this section, we implement the thermodynamic stability criteria
(\ref{eq:thermal_critera}) to the spin hydrodynamics. First, let
us briefly review the spin hydrodynamics in the canonical form. Besides
the energy momentum conservation, we also have the conservation equations
for the total angular momentum, i.e.
\begin{eqnarray}
\partial_{\lambda}J^{\lambda\mu\nu} & = & 0,\nonumber \\
\partial_{\mu}\Theta^{\mu\nu} & = & 0,\label{eq:conservation_eq_02}
\end{eqnarray}
where $J^{\lambda\mu\nu}$ and $\Theta^{\mu\nu}$ are the total angular
momentum tensor and energy momentum tensor in canonical form, respectively.
The constitutive equations of $J^{\lambda\mu\nu}$ and $\Theta^{\mu\nu}$
are
\begin{eqnarray}
\Theta^{\mu\nu} & = & (e+P)u^{\mu}u^{\nu}-Pg^{\mu\nu}+2q^{[\mu}u^{\nu]}+\phi^{\mu\nu}+\pi^{\mu\nu}-\Pi\Delta^{\mu\nu},\nonumber \\
J^{\lambda\mu\nu} & = & x^{\mu}\Theta^{\lambda\nu}-x^{\nu}\Theta^{\lambda\mu}+\Sigma^{\lambda\mu\nu},\label{eq:ThetaJ_Spin}
\end{eqnarray}
where $q^{\mu},\phi^{\mu\nu}$ are related to the spin and $\Sigma^{\lambda\mu\nu}$
is the spin tensor. In the following, we will limit our considerations
to the cases where (baryon) currents vanish and, therefore, the terms
for (baryon) number density do not contribute to constitutive relations
and thermodynamic relations.

Inserting Eq. (\ref{eq:ThetaJ_Spin}) into Eq. (\ref{eq:conservation_eq_02}),
yields
\begin{equation}
\partial_{\lambda}\Sigma^{\lambda\mu\nu}=-2\Theta^{[\mu\nu]}.\label{eq:spin_conservation_03}
\end{equation}
The spin tensor $\Sigma^{\lambda\mu\nu}$ is usually decomposed as
\citep{Hattori:2019lfp,Fukushima:2020ucl}
\begin{equation}
\Sigma^{\lambda\mu\nu}=u^{\lambda}S^{\mu\nu}+\sigma^{\lambda\mu\nu},
\end{equation}
where $S^{\mu\nu}$ is named as spin density and $\sigma^{\lambda\mu\nu}$
is perpendicular to the fluid velocity. We follow Ref. \citep{Fukushima:2020ucl}
to consider the power counting of the spin tensor, 
\begin{equation}
S^{\mu\nu}\sim\mathcal{O}(\partial^{0}),\quad\sigma^{\lambda\mu\nu}\sim\mathcal{O}(\partial^{1}).
\end{equation}
Analogy to charge chemical potential, one can introduce the spin chemical
potential $\omega^{\mu\nu}$, which modifies the thermodynamic relations
in the presence of spin density \citep{Hattori:2019lfp,Fukushima:2020ucl},
\begin{eqnarray}
e+P & = & Ts+\omega_{\mu\nu}S^{\mu\nu},\nonumber \\
de & = & Tds+\omega_{\mu\nu}dS^{\mu\nu},\nonumber \\
dP & = & sdT+S^{\mu\nu}d\omega_{\mu\nu}.\label{eq:ThermalRs}
\end{eqnarray}
The entropy current in Eq. (\ref{eq:entropy_current_01}) can also
be extended as

\begin{eqnarray}
s^{\mu} & = & su^{\mu}+\frac{1}{T}q^{\mu}-Q^{\mu}.\label{eq:Entropy_Spin}
\end{eqnarray}
The complete second order terms for the entropy current is complicated,
see e.g. Ref. \citep{Biswas:2023qsw}. For simplicity, we write down
the $Q^{\mu}$ analogy to Eq. (\ref{eq:Q_01}), 
\begin{equation}
Q^{\mu}=\frac{1}{2T}u^{\mu}(\chi_{q}q^{\nu}q_{\nu}+\chi_{\phi}\phi^{\alpha\beta}\phi_{\alpha\beta}+\chi_{\Pi}\Pi^{2}+\chi_{\pi}\pi^{\alpha\beta}\pi_{\alpha\beta}).
\end{equation}
From the second law of thermodynamics, we can get
\begin{eqnarray}
\tau_{q}\Delta^{\mu\nu}(u\cdot\partial)q_{\nu}+q^{\mu} & = & \lambda\left[u^{\rho}\partial_{\rho}u^{\mu}-T\Delta^{\mu\nu}\partial_{\nu}\frac{1}{T}-4\omega^{\mu\nu}u_{\nu}+\frac{1}{2}\chi_{q}T\partial_{\rho}\left(\frac{u^{\rho}}{T}\right)q_{\nu}\right],\nonumber \\
\tau_{\phi}\Delta^{\mu\alpha}\Delta^{\nu\beta}(u\cdot\partial)\phi_{\alpha\beta}+\phi^{\mu\nu} & = & 2\gamma_{s}\Delta^{\mu\alpha}\Delta^{\nu\beta}\left[\partial_{[\alpha}u_{\beta]}+2\omega_{\alpha\beta}-\frac{1}{2}\chi_{\phi}T\partial_{\rho}\left(\frac{u^{\rho}}{T}\right)\phi_{\alpha\beta}\right],\label{eq:q_phi_01}
\end{eqnarray}
with the transport coefficients, 
\begin{equation}
\tau_{q}=-\lambda\chi_{q},\quad\tau_{\phi}=2\chi_{\phi}\gamma_{s},\quad\lambda,\gamma_{s}>0.
\end{equation}
The equation for $\pi^{\mu\nu}$ and $\Pi$ are the same as Eq. (\ref{eq:Pi_pi}).
We notice that the terms proportional to $\chi_{q,}\chi_{\phi}$ on
the right-hand side of Eq. (\ref{eq:q_phi_01}) differs with the constitutive
equations for $q^{\mu}$ and $\phi^{\mu\nu}$ in the minimal causal
extended second order theory in Ref. \citep{Xie:2023gbo}. However,
these new terms proportional to $\chi_{q},\chi_{\phi}$ will not contribute
to the causality and stability conditions in linear mode analysis.

\subsection{Information current for spin hydrodynamics \label{subsec:InformationCSpin}}

Considering the small perturbations around thermodynamic equilibrium
$\varphi\rightarrow\varphi+\delta\varphi$, where $\varphi=(e,u^{\mu},\Pi,\pi^{\mu\nu},S^{\mu\nu},q^{\mu},\phi^{\mu\nu})$,
we can construct the information current $E^{\mu}$ for spin hydrodynamics.
According to the definition of $E^{\mu}$ in Eq. (\ref{eq:def_E}),
we next consider the conserved currents.

We note that different with Eq. (\ref{eq:Emu_01}), $u_{\nu}\delta\Theta^{\mu\nu}/T$
is no longer a conserved current in spin hydrodynamics due to the
nonvanishing antisymmetric part of $\delta\Theta^{\mu\nu}$. Recalling
that $u_{\mu}/T$ is a killing vector in thermodynamic equilibrium
state, i.e., $\partial_{(\mu}(u_{\nu)}/T)=0$, leading to the general
solutions for $u_{\mu}/T$ as \citep{Becattini:2012tc}
\begin{equation}
u_{\mu}/T=b_{\mu}+\varpi_{\mu\nu}x^{\nu},
\end{equation}
where $b_{\mu}$ and $\varpi_{\mu\nu}=-\varpi_{\nu\mu}$ are space-time
independent, and $\varpi_{\mu\nu}$ is named as the thermal vorticity
in spin hydrodynamics in the global equilibrium \citep{Hattori:2019lfp,Fukushima:2020ucl}.
Then, we find
\begin{equation}
\partial_{\mu}\left(\frac{u_{\nu}}{T}\delta\Theta^{\mu\nu}\right)=-\varpi_{\mu\nu}\delta\Theta^{[\mu\nu]},
\end{equation}
indicating that $u_{\nu}\delta\Theta^{\mu\nu}/T$ is not a conserved
current. According to Eq. (\ref{eq:spin_conservation_03}), we notice
that $\partial_{\mu}\delta\Sigma^{\mu\rho\sigma}=-2\delta\Theta^{[\mu\nu]}$,
and then construct a new conserved current $u_{\nu}\delta\Theta^{\mu\nu}/T-\frac{1}{2}\varpi_{\rho\sigma}\delta\Sigma^{\mu\rho\sigma}$,
which can also be written as  
\begin{equation}
\frac{u_{\nu}}{T}\delta\Theta^{\mu\nu}-\frac{1}{2}\varpi_{\rho\sigma}\delta\Sigma^{\mu\rho\sigma}=b_{\nu}\delta\Theta^{\mu\nu}-\frac{1}{2}\varpi_{\rho\sigma}\delta J^{\mu\rho\sigma}.\label{eq:ConservedCSpinHydro}
\end{equation}
The $b_{\nu}\delta\Theta^{\mu\nu}$ corresponds to energy and momentum
conservation. The $-\frac{1}{2}\varpi_{\rho\sigma}\delta J^{\mu\rho\sigma}$
comes from total angular momentum conservation. Interestingly, from
Eq. (\ref{eq:ConservedCSpinHydro}), the thermal vorticity $\varpi_{\mu\nu}$
plays a role like the chemical potential corresponding to the total
angular momentum. Numerous studies \citep{Becattini:2012tc,Becattini:2014yxa,Becattini:2018duy,Hattori:2019lfp,Fukushima:2020ucl,Hongo:2021ona}
prove that the thermal vorticity in the global equilibrium are proportional
to spin chemical potential, 
\begin{equation}
\varpi_{\rho\sigma}=\frac{2\omega_{\rho\sigma}}{T}.\label{eq:thermal_01}
\end{equation}

The independent currents in spin hydrodynamics are $u_{\nu}\delta\Theta^{\mu\nu}/T-\frac{1}{2}\varpi_{\rho\sigma}\delta\Sigma^{\mu\rho\sigma}$
and $\varpi_{\rho\sigma}\delta J^{\mu\rho\sigma}$. Recalling the
definition (\ref{eq:def_E}), we assume 
\begin{equation}
E^{\mu}=-\delta s^{\mu}+m_{1}\left(\frac{u_{\nu}}{T}\delta\Theta^{\mu\nu}-\frac{1}{2}\varpi_{\rho\sigma}\delta\Sigma^{\mu\rho\sigma}\right)+m_{2}\varpi_{\rho\sigma}\delta J^{\mu\rho\sigma},\label{eq:EmuAssump}
\end{equation}
with two constants $m_{1,2}$. Since the leading order of $E^{\mu}$
is $\mathcal{O}(\delta^{2})$ \citep{Gavassino:2021cli,Gavassino:2021kjm},
Eq. (\ref{eq:EmuAssump}) implies that 
\begin{equation}
\delta s^{\mu}=m_{1}\left(\frac{u_{\nu}}{T}\delta\Theta^{\mu\nu}-\frac{1}{2}\varpi_{\rho\sigma}\delta\Sigma^{\mu\rho\sigma}\right)+m_{2}\varpi_{\rho\sigma}\delta J^{\mu\rho\sigma},\label{eq:deltasa1a2_1}
\end{equation}
holds at order $\mathcal{O}(\delta)$. By contracting $u_{\mu}$ on
both sides of Eq. (\ref{eq:deltasa1a2_1}), we derive
\begin{equation}
\delta s=\frac{m_{1}}{T}\left(\delta e-\omega_{\rho\sigma}\delta S^{\rho\sigma}\right)+\frac{2m_{2}}{T}\omega_{\rho\sigma}\left(2x^{\rho}u_{\mu}\delta\Theta^{\mu\sigma}+\delta S^{\rho\sigma}\right),\label{eq:deltasa1a2_2}
\end{equation}
where the identity (\ref{eq:thermal_01}) is used. Comparison of Eq.
(\ref{eq:deltasa1a2_2}) with the thermodynamic relations (\ref{eq:ThermalRs})
yields $m_{1}=1$ and $m_{2}=0$, resulting in 
\begin{equation}
E^{\mu}=-\delta s^{\mu}+\frac{u_{\nu}}{T}\delta\Theta^{\mu\nu}-\frac{1}{T}\omega_{\rho\sigma}\delta\Sigma^{\mu\rho\sigma}.\label{eq:EmuSpinHydro}
\end{equation}
Following the same strategy as in Sec. \ref{sec:MIS}, we will choose
the rest frame of the fluid without rotation and assume the irrotational
system reaches the thermodynamic equilibrium, 
\begin{equation}
\{q^{\mu},\phi^{\mu\nu},\omega^{\mu\nu},S^{\mu\nu}\}=0.\label{eq:equalibirum}
\end{equation}

The perturbation $\delta s$ in Eq. (\ref{eq:delta_s_01}) becomes
\begin{equation}
\delta s=\frac{1}{T}\delta e-\frac{1}{2T}\frac{c_{s}^{2}}{e+P}(\delta e)^{2}-\frac{1}{2T}\delta\omega_{\alpha\beta}\delta S^{\alpha\beta}+O(\delta^{3}).
\end{equation}
With the above results and Eqs. (\ref{eq:ThetaJ_Spin}), (\ref{eq:ThermalRs}),
and (\ref{eq:Entropy_Spin}), the information current can be expressed
as 
\begin{eqnarray}
E^{\mu} & = & -\delta s^{\mu}+\frac{1}{T}u_{\nu}\delta\Theta^{\mu\nu}-\frac{1}{T}\omega_{\rho\sigma}\delta\Sigma^{\mu\rho\sigma}\nonumber \\
 & = & \frac{1}{2T}\frac{c_{s}^{2}}{e+P}(\delta e)^{2}u^{\mu}+\frac{c_{s}^{2}}{T}\delta e\delta u^{\mu}+\frac{c_{s}^{2}}{T(e+P)}\delta e\delta q^{\mu}+\frac{1}{2T}\delta\omega_{\alpha\beta}\delta S^{\alpha\beta}u^{\mu}\nonumber \\
 &  & -\frac{1}{2T}(e+P)u^{\mu}\delta u_{\nu}\delta u^{\nu}+\frac{1}{T}\delta u_{\nu}\delta q^{\nu}u^{\mu}-\frac{1}{T}\delta u_{\nu}\delta\phi^{\mu\nu}-\frac{1}{T}\delta u_{\nu}\delta\pi^{\mu\nu}+\frac{1}{T}\delta\Pi\delta u^{\mu}\nonumber \\
 &  & +\frac{1}{2T}u^{\mu}(\chi_{q}\delta q^{\nu}\delta q_{\nu}+\chi_{\phi}\delta\phi^{\alpha\beta}\delta\phi_{\alpha\beta}+\chi_{\Pi}\delta\Pi\delta\Pi+\chi_{\pi}\delta\pi^{\alpha\beta}\delta\pi_{\alpha\beta}),\label{eq:EmuSpinHdro01}
\end{eqnarray}
where we have used
\begin{eqnarray}
u_{\mu}\delta q^{\mu} & = & -\delta u_{\mu}\delta q^{\mu},\nonumber \\
u_{\nu}\delta\phi^{\mu\nu} & = & -\delta u_{\nu}\delta\phi^{\mu\nu}.
\end{eqnarray}
As a cross-check, we derive Eq. (\ref{eq:EmuSpinHdro01}) by using
a different approach shown in Appendix \ref{sec:An-alternative-approach}.

Again, let us take $u^{\mu}=(1,0)$. For arbitrary $n^{\mu}$ with
$\ensuremath{n_{0}>0}$ and $n^{\mu}n_{\mu}=1$, we can get 
\begin{eqnarray}
\frac{2n_{0}TE^{\mu}n_{\mu}}{e+P} & = & \frac{n_{0}^{2}\tau_{\pi}}{\eta(e+P)}\sum_{i<j}\left[\delta\pi^{ij}-\frac{1}{n_{0}\chi_{\pi}}n_{(i}\delta u_{j)}\right]^{2}\nonumber \\
 &  & +\frac{n_{0}^{2}\tau_{\pi}}{\eta(e+P)}\left[\delta\pi^{11}+\frac{1}{2}\delta\pi^{22}+\frac{1}{2n_{0}\chi_{\pi}}(n_{3}\delta u_{3}-n_{1}\delta u_{1})\right]^{2}\nonumber \\
 &  & +\frac{3n_{0}^{2}\tau_{\pi}}{4\eta(e+P)}\left[\delta\pi^{22}+\frac{1}{3n_{0}\chi_{\pi}}(n_{3}\delta u_{3}+n_{1}\delta u_{1}-2n_{2}\delta u_{2})\right]^{2}\nonumber \\
 &  & +\frac{n_{0}^{2}\tau_{q}}{\lambda(e+P)}\sum_{i}\left[\delta q^{i}-\frac{1}{n_{0}\chi_{q}}\left(\frac{c_{s}^{2}}{e+P}\delta en_{i}+n_{0}\delta u_{i}\right)\right]^{2}\nonumber \\
 &  & +\frac{n_{0}^{2}\tau_{\phi}}{\gamma_{s}(e+P)}\sum_{i<j}\left(\delta\phi^{ij}-\frac{1}{n_{0}\chi_{\phi}}n_{[i}\delta u_{j]}\right)^{2}+\frac{n_{0}^{2}}{e+P}\delta\omega_{\alpha\beta}\delta S^{\alpha\beta}\nonumber \\
 &  & +\sum_{i=6}^{10}a_{i}(\delta A_{i})^{2}+O(\delta^{3}),\label{eq:E_n_01}
\end{eqnarray}
where the expressions for $a_{i}$ and $\delta A_{i}$ are presented
in Appendix \ref{sec:Expressions_aiAi}. Next, we analyze the thermodynamic
stability in two cases: with and without viscous tensors, $\pi^{\mu\nu}$
and $\Pi\Delta^{\mu\nu}$. The main reason is as follows. In the previous
study by some of us \citep{Xie:2023gbo}, we find that there exist
zero modes in the linear mode analysis for the spin hydrodynamics
with vanishing viscous tensors. Such zero modes disappear once we
turn on the finite viscous tensors. It is questionable whether the
spin hydrodynamics can be stable and causal with vanishing viscous
tensors. Therefore, it is necessary to study the thermodynamic stability
with and without viscous tensors separately.

\subsection{Case I: With vanishing viscous tensors \label{subsec:CaseI}}

By simply setting $\delta\pi^{\mu\nu}$ and $\delta\Pi$ to zero in
Eq. (\ref{eq:E_n_01}), we find that the sufficient and necessary
conditions for thermodynamic stability (\ref{eq:thermal_critera})
are 
\begin{eqnarray}
c_{s}^{2},\gamma_{s},\lambda,\tau_{\phi},\tau_{q},\delta\omega_{\alpha\beta}\delta S^{\alpha\beta} & > & 0,\nonumber \\
1-\frac{\lambda}{\tau_{q}(e+P)}-\frac{\gamma_{s}}{\tau_{\phi}(e+P)} & > & 0,\nonumber \\
1-c_{s}^{2}-\frac{(3c_{s}^{2}+1)\lambda}{\tau_{q}(e+P)} & > & 0.\label{eq:conditions_wo_viscous}
\end{eqnarray}
The last two inequalities can be rewritten as 
\begin{eqnarray}
0<\frac{2\gamma^{\prime}\tau_{q}}{(2\tau_{q}-\lambda^{\prime})\tau_{\phi}} & < & 1,\nonumber \\
0<\frac{c_{s}^{2}(2\tau_{q}+3\lambda^{\prime})}{2\tau_{q}-\lambda^{\prime}} & < & 1,\label{eq:Conditions_01}
\end{eqnarray}
where 
\begin{equation}
\lambda^{\prime}=\frac{2\lambda}{e+P},\ \gamma^{\prime}=\frac{\gamma_{s}}{e+P}.\label{eq:LinearNotations0}
\end{equation}
We find that the conditions (\ref{eq:Conditions_01}) are exactly
the same as the causality conditions derived by linear mode analysis
\citep{Xie:2023gbo}.

The stability conditions from linear mode analysis are given by \citep{Xie:2023gbo}
\begin{eqnarray}
c_{s}^{2},\gamma_{s},\lambda,\tau_{\phi},\tau_{q},\chi_{s},-\chi_{b} & > & 0,\nonumber \\
2\tau_{q}-\lambda^{\prime} & > & 0,\nonumber \\
\chi_{e}^{0i} & = & 0,\label{eq:LinearStable_qphi}
\end{eqnarray}
where $\chi_{e}^{\mu\nu}$ and $\chi_{b},\chi_{s}$ are the spin susceptibilities
with respect to $e$ and $S^{0i},S^{ij}$, i.e.
\begin{eqnarray}
\delta\omega^{0i} & = & \chi_{e}^{0i}\delta e+\chi_{b}\delta S^{0i},\nonumber \\
\delta\omega^{ij} & = & \chi_{e}^{ij}\delta e+\chi_{s}\delta S^{ij}.\label{eq:spin_susp_01}
\end{eqnarray}
The inequality $2\tau_{q}>\lambda^{\prime}$ can be directly derived
from the thermodynamic stability conditions (\ref{eq:conditions_wo_viscous}).

With the parametrization (\ref{eq:spin_susp_01}), the inequalities
$\chi_{b}<0$ and $\chi_{s}>0$ are necessary conditions for $\delta\omega_{\alpha\beta}\delta S^{\alpha\beta}>0$
in Eq. (\ref{eq:conditions_wo_viscous}). However, $\chi_{e}^{0i}=0$
does not arise immediately from the thermodynamic stability conditions.
In fact, the spin susceptibility $\chi_{e}^{\mu\nu}$ introduced in
Eq. (\ref{eq:spin_susp_01}) is a high order correction in our setup.
Let us consider the equations of state,
\begin{equation}
e=e(T,\omega^{\mu\nu}),\quad S^{\mu\nu}=S^{\mu\nu}(T,\omega^{\mu\nu}).
\end{equation}
For simplicity, let us focus on $S^{xy}$ and $\omega^{xy}$, and
assume other components of $S^{\mu\nu}$ and $\omega^{\mu\nu}$ are
vanishing. Since the $\omega^{xy}\sim\mathcal{O}(\partial^{1})$ is
the quantum correction to the thermodynamic variables, the equations
of state can be expressed as power series of $\omega^{xy}$ based
on symmetry considerations\footnote{Here, we assume the absence of characteristic or external tensors.
In other words, the system is considered ``isotropic.'' Clearly,
this assumption implies $\chi_{e}^{\mu\nu}\sim\omega^{\mu\nu}$ by
considering the antisymmetric tensor structure of $\chi_{e}^{\mu\nu}$.
If this assumption does not hold, then $\chi_{e}^{\mu\nu}$ may be
nonzero even if $\omega^{\mu\nu}=0$.}, 
\begin{equation}
\left(\begin{array}{c}
\delta e\\
\delta S^{xy}
\end{array}\right)=\left(\begin{array}{cc}
a_{11}T^{3} & a_{12}\omega^{xy}T^{2}\\
a_{21}\omega^{xy}T & a_{22}T^{2}
\end{array}\right)\left(\begin{array}{c}
\delta T\\
\delta\omega^{xy}
\end{array}\right)+\mathcal{O}(\omega_{xy}^{2}\delta\omega^{xy},\omega_{xy}^{2}\delta T),\label{eq:DeltapS}
\end{equation}
where $a_{ij}$ are dimensionless constants and $a_{11},a_{22}\neq0$.
The inverse of Eq. (\ref{eq:DeltapS}) gives 
\begin{equation}
\left(\begin{array}{c}
\delta T\\
\delta\omega^{xy}
\end{array}\right)=\frac{1}{a_{11}a_{22}T^{4}}\left(\begin{array}{cc}
a_{22}T & -a_{12}\omega^{xy}T\\
-a_{21}\omega^{xy} & a_{11}T^{2}
\end{array}\right)\left(\begin{array}{c}
\delta e\\
\delta S^{xy}
\end{array}\right)+\mathcal{O}(\omega_{xy}^{2}\delta e,\omega_{xy}^{2}\delta S^{\mu\nu}).\label{eq:DeltaTomega}
\end{equation}
We find that $\chi_{e}^{xy}\propto\omega^{xy}$. When the system reaches
irrotational equilibrium state shown in Eq. (\ref{eq:equalibirum}),
$\chi_{e}^{xy}\propto\delta\omega^{xy}$, therefore $\chi_{e}^{xy}\delta e\sim\mathcal{O}(\delta^{2})$
are high order corrections. While $\chi_{s}\sim1/(a_{22}T^{2})\sim\mathcal{O}(\delta^{0})$
can survive. Hence, the condition $\chi_{e}^{\mu\nu}=\mathcal{O}(\delta)$
does not arise from stability demand but rather from our choice of
an irrotational background.

Taking the parameterization (\ref{eq:spin_susp_01}) with $\chi_{e}^{\mu\nu}=\mathcal{O}(\delta)$,
the inequalities $\chi_{b}<0$ and $\chi_{s}>0$ now become equivalent
to $\delta\omega_{\alpha\beta}\delta S^{\alpha\beta}>0$. Consequently,
in the case of $\Pi,\pi^{\mu\nu}=0$, the thermodynamic stability
conditions align with the stability and causality conditions derived
from linear mode analysis in Ref. \citep{Xie:2023gbo}. It also indicates
that the zero modes in the dispersion relations appeared in linear
mode analysis \citep{Xie:2023gbo} will not lead to instabilities.

\subsection{Case II: With finite viscous tensors \label{subsec:Case-II}}

Let us consider the full form of $E^{\mu}$ shown in Eq. (\ref{eq:EmuSpinHdro01}).
Imposing the thermodynamic stability conditions (\ref{eq:thermal_critera})
yields
\begin{eqnarray}
c_{s}^{2},\lambda,\gamma_{s},\eta,\zeta,\tau_{q},\tau_{\phi},\tau_{\pi},\tau_{\Pi},-\chi_{b},\chi_{s} & > & 0,\label{eq:ThermalR_1}\\
1-\frac{\lambda^{\prime}}{2\tau_{q}}-\frac{4\gamma_{\perp}}{3\tau_{\pi}}-\frac{1}{3\tau_{\Pi}}(3\gamma_{\|}-4\gamma_{\perp}) & > & 0,\label{eq:ThermalR_2}\\
1-\frac{\lambda^{\prime}}{2\tau_{q}}-\frac{\gamma_{\perp}}{\tau_{\pi}}-\frac{\gamma^{\prime}}{\tau_{\phi}} & > & 0,\label{eq:ThermalR_3}\\
1-c_{s}^{2}-\frac{(1+3c_{s}^{2})\lambda^{\prime}}{2\tau_{q}}-\frac{(2\tau_{q}-c_{s}^{2}\lambda^{\prime})[4\gamma_{\perp}\tau_{\Pi}+\tau_{\pi}(3\gamma_{\|}-4\gamma_{\perp})]}{6\tau_{q}\tau_{\pi}\tau_{\Pi}} & > & 0,\label{eq:ThermalR_4}\\
2-c_{s}^{2}-\frac{(2+3c_{s}^{2})\lambda^{\prime}}{2\tau_{q}}-\frac{4\gamma_{\perp}\tau_{\Pi}+\tau_{\pi}(3\gamma_{\|}-4\gamma_{\perp})}{3\tau_{\pi}\tau_{\Pi}} & > & 0,\label{eq:ThermalR_5}
\end{eqnarray}
where we have used the parametrization (\ref{eq:spin_susp_01}) and
the shorthand notations (\ref{eq:LinearNotations0}) and 
\begin{equation}
\gamma_{\perp}=\frac{\eta}{e+P},\ \gamma_{\|}=\frac{\frac{4}{3}\eta+\zeta}{e+P}.\label{eq:LinearNotations1}
\end{equation}

We now compare these conditions (\ref{eq:ThermalR_1})-(\ref{eq:ThermalR_5})
to those derived from linear mode analysis \citep{Xie:2023gbo}. The
causality conditions in linear mode analysis are given by
\begin{eqnarray}
0<\frac{2\tau_{q}(\gamma^{\prime}\tau_{\pi}+\gamma_{\perp}\tau_{\phi})}{(2\tau_{q}-\lambda^{\prime})\tau_{\pi}\tau_{\phi}} & < & 1,\label{eq:LinearCausal_1}\\
0<\frac{b_{1}^{1/2}\pm(b_{1}-b_{2})^{1/2}}{6(2\tau_{q}-\lambda^{\prime})\tau_{\pi}\tau_{\Pi}} & < & 1,\label{eq:LinearCausal_2}
\end{eqnarray}
where $b_{1,2}$ are defined as 
\begin{eqnarray}
b_{1}^{1/2} & = & 8\gamma_{\perp}\tau_{q}\tau_{\Pi}+\tau_{\pi}[2\tau_{q}(3\gamma_{\|}-4\gamma_{\perp})+3\tau_{\Pi}c_{s}^{2}(3\lambda^{\prime}+2\tau_{q})],\nonumber \\
b_{2} & = & 12c_{s}^{2}\lambda^{\prime}(2\tau_{q}-\lambda^{\prime})\tau_{\pi}\tau_{\Pi}[\tau_{\pi}(3\gamma_{\|}-4\gamma_{\perp})+4\gamma_{\perp}\tau_{\Pi}].
\end{eqnarray}
It is straightforward to show that the inequality (\ref{eq:LinearCausal_1})
can be derived from inequalities (\ref{eq:ThermalR_1}) and (\ref{eq:ThermalR_3}).
Similarly, one can derive (\ref{eq:LinearCausal_2}) by using inequalities
(\ref{eq:ThermalR_1}) and (\ref{eq:ThermalR_5}). We then conclude
that the causality in linear mode analysis is ensured by thermodynamic
stability conditions.

The stability conditions derived by linear mode analysis are \citep{Xie:2023gbo}
\begin{eqnarray}
c_{s}^{2},\lambda,\gamma_{s},\eta,\zeta,\tau_{q},\tau_{\phi},\tau_{\pi},\tau_{\Pi},-\chi_{b},\chi_{s} & > & 0,\label{eq:LinearStable_1}\\
2\tau_{q}-\lambda^{\prime} & > & 0,\label{eq:LinearStable_2}\\
b_{1}>b_{2} & > & 0,\label{eq:LinearStable_3}\\
\frac{c_{2}}{c_{3}} & > & 0,\label{eq:LinearStable_4}
\end{eqnarray}
where the definitions of $c_{2,3}$ are presented in Appendix \ref{sec:Expressionsbici}.
After performing the calculations detailed in Appendix \ref{sec:Stability},
we show that the inequalities (\ref{eq:LinearStable_3}), (\ref{eq:LinearStable_4})
can be derived from (\ref{eq:LinearStable_1}), (\ref{eq:LinearStable_2}).
Consequently, the independent stability conditions in linear mode
analysis reduce to Eqs. (\ref{eq:LinearStable_1}), (\ref{eq:LinearStable_2}).
It is worth noting that the inequality (\ref{eq:LinearStable_1})
aligns precisely with inequality (\ref{eq:ThermalR_1}) under the
parametrization (\ref{eq:spin_susp_01}), while inequality (\ref{eq:LinearStable_2})
can be derived from either inequality (\ref{eq:ThermalR_2}) or (\ref{eq:ThermalR_3}).

Our results reveal that the stability and causality conditions derived
in linear mode analysis can indeed be derived from thermodynamic stability
conditions. However, the reverse does not hold in the current case.
For instance, the inequality (\ref{eq:ThermalR_2}) cannot be derived
from the causality and stability conditions identified in linear mode
analysis. Therefore, unlike the scenarios discussed in Secs. \ref{sec:MIS}
and \ref{subsec:CaseI}, the thermodynamic stability conditions for
spin hydrodynamics involving nonvanishing components $q^{\mu}$, $\phi^{\mu\nu}$,
$\Pi$, and $\pi^{\mu\nu}$ are more stringent than those derived
from linear mode analysis.

Let us discuss the above observation. A dissipative process is called
real or on shell if it satisfies the equations of motion, otherwise,
it is called virtual or off shell. Linear mode analysis solely considers
real processes, whereas thermodynamic stability analysis encompasses
both real and virtual processes \citep{Gavassino:2021kjm,Gavassino:2024vyu}.
If there are no virtual processes, meaning all forms of perturbations
are allowed, then the conditions derived from thermodynamic stability
analysis and linear mode analysis coincide, as the cases in Secs.
\ref{sec:MIS} and \ref{subsec:CaseI}. However, in the presence of
virtual processes, additional conditions emerge from thermodynamic
stability analysis and are invisible in linear mode analysis. Consequently,
the thermodynamic stability are more stringent compared to linear-mode
stability. This implies that the thermodynamic stability analysis
for spin hydrodynamics with viscous tensors may involve virtual processes
that are not allowed by linearized hydrodynamic equations. A systematic
verification of this statement is left for our future work.

In Ref. \citep{Xie:2023gbo}, it was found that the conditions derived
from linear mode analysis might be necessary but are not sufficient
to ensure stability. In contrast, the thermodynamic stability criteria
(\ref{eq:thermal_critera}) are both necessary and sufficient for
ensuring stability. The reasoning is as follows.

Clearly, the thermodynamic stability criteria (\ref{eq:thermal_critera})
are necessary to uphold the fundamental laws of stability, specifically
the second law of thermodynamics and the principle of maximizing total
entropy in the equilibrium state. On the other hand, the functional
$E[\delta\varphi]$ defined in Eq. (\ref{eq:E_def}) is positive definite
and nonincreasing in time when the criteria (\ref{eq:thermal_critera})
are fulfilled. Then $E[\delta\varphi]$ can be interpreted as a Lyapunov
functional, which is sufficient to guarantee the stability of the
corresponding linearized hydrodynamic equations \citep{lasalle1961stability,Gavassino:2021kjm,Gavassino:2023odx}.
Therefore, we argue that the unstable modes identified in Ref. \citep{Xie:2023gbo}
would disappear if we adopt the conditions from thermodynamic stability
(\ref{eq:ThermalR_1})-(\ref{eq:ThermalR_5}). A rigorous proof of
this assertion will require more general discussions on the structure
of linearized hydrodynamic equations and will be presented elsewhere.

\section{Conclusion \label{sec:Conclusion}}

In this work, we have applied thermodynamic stability analysis to
derive the stability and causality conditions for conventional relativistic
viscous hydrodynamics and spin hydrodynamics.

As a test, we first derived the thermodynamic stability conditions
in Eq. (\ref{eq:MISC2}) for second-order relativistic viscous hydrodynamics
without (baryon) currents and heat currents. We found that these conditions
are consistent with those derived from linear mode analysis in Refs.
\citep{Denicol:2008ha,Pu:2009fj,Pu:2011vr}.

We next studied the thermodynamic stability of minimal causal extended
second-order spin hydrodynamics in canonical form, both with and without
viscous tensors. In the absence of viscous tensors, the constraints
derived from thermodynamic stability analysis exactly match those
obtained from linear mode analysis. This indicates that the zero modes
found in the linear mode analysis will not affect the causality and
stability of the spin hydrodynamics in this case.

As another important observation, we also note that the inequality
$\delta\omega_{\alpha\beta}\delta S^{\alpha\beta}>0$ in Eq. (\ref{eq:conditions_wo_viscous})
can be satisfied by adopting physical equations of state. The spin
susceptibilities with respect to energy density, $\chi_{e}^{\mu\nu}$,
are found to be $\sim\mathcal{O}(\delta)$ and therefore can be neglected
in the current setup. This finding could help us understand the unstable
modes identified in Ref. \citep{Xie:2023gbo} when the asymptotic
stability conditions are met in the linear modes analysis.

We then derive the thermodynamic stability conditions in Eqs. (\ref{eq:ThermalR_1})-(\ref{eq:ThermalR_5})
for spin hydrodynamics in the presence of viscous tensors. These conditions
are consistent with the causality conditions derived from linear mode
analysis and are more stringent than the stability conditions found
in linear mode analysis. Our studies suggest that the conditions derived
from thermodynamic stability analysis can guarantee both causality
and stability in linear mode analysis.

In the current studies, we have only considered irrotational spin
hydrodynamics. The inclusion of a rotating background will affect
the analysis, as noted in Ref. \citep{Shokri:2023rpp}, and should
be studied systematically in future work.
\begin{acknowledgments}
We thank Lorenzo Gavassino for explaining the differences between
thermodynamic stability analysis and linear mode analysis and Masoud
Shokri for fruitful discussions on the information current for spin
hydrodynamics. This work is supported in part by the National Key
Research and Development Program of China under Contract No. 2022YFA1605500,
by the Chinese Academy of Sciences (CAS) under Grant No. YSBR-088
and by National Nature Science Foundation of China (NSFC) under Grants
No. 12075235 and No.12135011.
\end{acknowledgments}

\appendix

\section{Another approach to derive the information current for spin hydrodynamics
\label{sec:An-alternative-approach}}

Here we employ the method used in Ref. \citep{Gavassino:2023qnw}
(see also the Supplemental Material of Ref. \citep{Gavassino:2021kjm})
to derive the information current (\ref{eq:EmuSpinHdro01}) for spin
hydrodynamics. This method is based on the fact that the function
$\Psi$, defined in Eq. (\ref{eq:Phi}), should be maximized in the
equilibrium state. We now introduce $\theta$ to characterize a smooth
one-parameter family of solutions to hydrodynamic equations, where
only $\theta=0$ corresponds to the equilibrium state. Then $\Psi=\Psi(\theta)$
is a function of $\theta$. Since $\Psi$ is maximized in the equilibrium
state, we have 
\begin{equation}
\dot{\Psi}(0)=0,\quad\ddot{\Psi}(0)\leq0,\label{eq:DotPhi}
\end{equation}
where the dot represents the derivative with respect to $\theta$.
Given an arbitrary three-dimensional spacelike Cauchy surface $\Sigma$
with the future-directed and timelike normal unit vector $n^{\mu}$,
we can express $\Psi$ as $\Psi=\int_{\Sigma}d\Sigma n_{\mu}\psi^{\mu}$,
with the current $\psi^{\mu}=\psi^{\mu}(\theta)$ given by 
\begin{equation}
\psi^{\mu}=s^{\mu}+\sum_{a}\alpha^{a}J^{a,\mu}.
\end{equation}
Due to arbitrariness of the Cauchy surface $\Sigma$, Eq. (\ref{eq:DotPhi})
implies that 
\begin{equation}
\dot{\psi}^{\mu}(0)=0,\label{eq:Dotphimu}
\end{equation}
and $\ddot{\psi}^{\mu}(0)$ is past-directed and nonspacelike. For
small $\theta$, the information current $E^{\mu}$ can be derived
through \citep{Gavassino:2021kjm,Gavassino:2023qnw}
\begin{equation}
E^{\mu}=-\frac{1}{2}\theta^{2}\ddot{\psi}^{\mu}(0).\label{eq:Emu2}
\end{equation}

To calculate the the information current $E^{\mu}$ using Eq. (\ref{eq:Emu2}),
let us first construct the current $\psi^{\mu}$. According to the
discussion in Sec. \ref{subsec:InformationCSpin}, there are two independent
conserved currents, 
\begin{equation}
\kappa_{\nu}\Theta^{\mu\nu}+\frac{1}{2}\partial_{[\rho}\kappa_{\sigma]}\Sigma^{\mu\rho\sigma},\quad\xi_{\rho\sigma}J^{\mu\rho\sigma},
\end{equation}
where $\kappa^{\mu}$ is a killing vector and $\xi_{\rho\sigma}$
is an antisymmetric constant tensor. The general form for $\psi^{\mu}$
is 
\begin{equation}
\psi^{\mu}=s^{\mu}-\kappa_{\nu}\Theta^{\mu\nu}-\frac{1}{2}\partial_{[\rho}\kappa_{\sigma]}\Sigma^{\mu\rho\sigma}-\xi_{\rho\sigma}J^{\mu\rho\sigma}.\label{eq:phimuGeneral}
\end{equation}
By introducing another killing vector 
\begin{equation}
\beta_{\nu}=\kappa_{\nu}+2\xi_{\rho\nu}x^{\rho},
\end{equation}
the expression (\ref{eq:phimuGeneral}) can be equivalently written
as 
\begin{equation}
\psi^{\mu}=s^{\mu}-\beta_{\nu}\Theta^{\mu\nu}-\frac{1}{2}\partial_{[\rho}\beta_{\sigma]}\Sigma^{\mu\rho\sigma}.
\end{equation}
Substituting the constitutive equations (\ref{eq:ThetaJ_Spin}) into
it, we obtain  
\begin{eqnarray}
\psi^{\mu} & = & \left[s-(e+P+\Pi)\beta_{\nu}u^{\nu}-\frac{1}{2}\partial_{[\rho}\beta_{\sigma]}S^{\rho\sigma}+q^{\nu}\beta_{\nu}-\mathcal{K}\right]u^{\mu}\nonumber \\
 &  & +(P+\Pi)\beta^{\mu}-q^{\mu}(\beta_{\nu}u^{\nu}-\frac{1}{T})-(\phi^{\mu\nu}+\pi^{\mu\nu})\beta_{\nu},
\end{eqnarray}
where 
\begin{equation}
\mathcal{K}=\frac{1}{2T}(\chi_{q}q^{\nu}q_{\nu}+\chi_{\phi}\phi^{\alpha\beta}\phi_{\alpha\beta}+\chi_{\Pi}\Pi^{2}+\chi_{\pi}\pi^{\alpha\beta}\pi_{\alpha\beta}).
\end{equation}

The next step is to impose the constraint (\ref{eq:Dotphimu}) on
$\psi^{\mu}$. We find 
\begin{eqnarray}
\dot{\psi}^{\mu} & = & \left[\dot{s}-(\dot{e}+\dot{P}+\dot{\Pi})\beta_{\nu}u^{\nu}-(e+P+\Pi)\beta_{\nu}\dot{u}^{\nu}-\frac{1}{2}\partial_{[\rho}\beta_{\sigma]}\dot{S}^{\rho\sigma}+\dot{q}^{\nu}\beta_{\nu}-\dot{\mathcal{K}}\right]u^{\mu}\nonumber \\
 &  & +\left[s-(e+P+\Pi)\beta_{\nu}u^{\nu}-\frac{1}{2}\partial_{[\rho}\beta_{\sigma]}S^{\rho\sigma}+q^{\nu}\beta_{\nu}-\mathcal{K}\right]\dot{u}^{\mu}\nonumber \\
 &  & +(\dot{P}+\dot{\Pi})\beta^{\mu}-\dot{q}^{\mu}(\beta_{\nu}u^{\nu}-\frac{1}{T})-q^{\mu}(\beta_{\nu}\dot{u}^{\nu}+\frac{1}{T^{2}}\dot{T})-(\dot{\phi}^{\mu\nu}+\dot{\pi}^{\mu\nu})\beta_{\nu}.
\end{eqnarray}
Note that here $u^{\mu}$ and $\dot{u}^{\mu}$ are independent, and
this is true for other variables. The constraint (\ref{eq:Dotphimu})
demands 
\begin{equation}
\frac{u_{\nu}}{T}=\beta_{\nu},\quad\frac{2}{T}\omega_{\rho\sigma}=-\partial_{[\rho}\beta_{\sigma]},\quad\Pi,q^{\mu},\phi^{\mu\nu},\pi^{\mu\nu}=0,\label{eq:EquilibriumC0}
\end{equation}
in the equilibrium state. These conditions are exactly the same as
those from entropy current analysis \citep{Hattori:2019lfp,Fukushima:2020ucl}.

With the equilibrium conditions (\ref{eq:EquilibriumC0}), we can
get 
\begin{eqnarray}
u_{\nu}\ddot{u}^{\nu} & = & \dot{u}_{\nu}\dot{u}^{\nu},\quad u_{\nu}\dot{q}^{\nu}=0,\quad u_{\nu}\ddot{q}^{\nu}=-2\dot{u}_{\nu}\dot{q}^{\nu},\nonumber \\
\ddot{\phi}^{\mu\nu}u_{\nu} & = & -2\dot{\phi}^{\mu\nu}\dot{u}_{\nu},\quad\ddot{\pi}^{\mu\nu}u_{\nu}=-2\dot{\pi}^{\mu\nu}\dot{u}_{\nu}.\label{eq:ids01}
\end{eqnarray}
The thermodynamic relations (\ref{eq:ThermalRs}) give 
\begin{equation}
\ddot{e}=T\ddot{s}+\omega_{\rho\sigma}\ddot{S}^{\rho\sigma}+\dot{T}\dot{s}+\dot{\omega}_{\rho\sigma}\dot{S}^{\rho\sigma}.\label{eq:ids02}
\end{equation}
With the help of these identities (\ref{eq:ids01}), (\ref{eq:ids02}),
we derive 
\begin{eqnarray}
\ddot{\psi}^{\mu}(0) & = & -\left[\frac{1}{T}\dot{T}\dot{s}+\frac{1}{T}\dot{\omega}_{\rho\sigma}\dot{S}^{\rho\sigma}-\frac{1}{T}(e+P)\dot{u}_{\nu}\dot{u}^{\nu}+\frac{2}{T}\dot{u}_{\nu}\dot{q}^{\nu}\right]u^{\mu}\nonumber \\
 &  & -\frac{1}{T}(\chi_{q}\dot{q}^{\nu}\dot{q}_{\nu}+\chi_{\phi}\dot{\phi}^{\alpha\beta}\dot{\phi}_{\alpha\beta}+\chi_{\Pi}\dot{\Pi}^{2}+\chi_{\pi}\dot{\pi}^{\alpha\beta}\dot{\pi}_{\alpha\beta})u^{\mu}\nonumber \\
 &  & -\frac{2}{T}(\dot{P}+\dot{\Pi})\dot{u}^{\mu}-\frac{2}{T^{2}}\dot{q}^{\mu}\dot{T}+\frac{2}{T}\dot{\phi}^{\mu\nu}\dot{u}_{\nu}+\frac{2}{T}\dot{\pi}^{\mu\nu}\dot{u}_{\nu}.
\end{eqnarray}
Notice that, for small $\theta$, the quantity $\theta\dot{\varphi}$
represent the small perturbation around the equilibrium state, i.e.
\begin{equation}
\delta\varphi=\theta\dot{\varphi},
\end{equation}
where $\varphi$ stands for the hydrodynamic variables $T,s,\Pi,u^{\mu},q^{\mu}$,
etc. Hence, the information current from Eq. (\ref{eq:Emu2}) can
be expressed as 
\begin{eqnarray}
E^{\mu} & = & \frac{1}{2}\left[\frac{1}{T}\delta T\delta s+\frac{1}{T}\delta\omega_{\rho\sigma}\delta S^{\rho\sigma}-\frac{1}{T}(e+P)\delta u_{\nu}\delta u^{\nu}+\frac{2}{T}\delta u_{\nu}\delta q^{\nu}\right]u^{\mu}\nonumber \\
 &  & +\frac{1}{2T}(\chi_{q}\delta q^{\nu}\delta q_{\nu}+\chi_{\phi}\delta\phi^{\alpha\beta}\delta\phi_{\alpha\beta}+\chi_{\Pi}\delta\Pi\delta\Pi+\chi_{\pi}\delta\pi^{\alpha\beta}\delta\pi_{\alpha\beta})u^{\mu}\nonumber \\
 &  & +\frac{1}{T}(\delta P+\delta\Pi)\delta u^{\mu}+\frac{1}{T^{2}}\delta q^{\mu}\delta T-\frac{1}{T}\delta\phi^{\mu\nu}\delta u_{\nu}-\frac{1}{T}\delta\pi^{\mu\nu}\delta u_{\nu}+\mathcal{O}(\delta^{3}).\label{eq:EmuGeneral01}
\end{eqnarray}
The formula (\ref{eq:EmuGeneral01}) works for both rotational and
irrotational background.

In an irrotational background where $\omega^{\mu\nu},S^{\mu\nu}=0$,
we have 
\begin{eqnarray}
\delta s & = & \frac{1}{T}\delta e+\mathcal{O}(\delta^{2}),\nonumber \\
\delta P & = & c_{s}^{2}\delta e+\mathcal{O}(\delta^{2}),\nonumber \\
\delta T & = & \frac{c_{s}^{2}T}{e+P}\delta e+\mathcal{O}(\delta^{2}).\label{eq:Irrotational_spt}
\end{eqnarray}
Plugging Eq. (\ref{eq:Irrotational_spt}) into Eq. (\ref{eq:EmuGeneral01}),
we obtain the same information current as Eq. (\ref{eq:EmuSpinHdro01}).

\section{Expressions for $a_{i}$ and $\delta A_{i}$ in Eqs. (\ref{eq:En_01}),
(\ref{eq:E_n_01}) \label{sec:Expressions_aiAi}}

Here, we present the expressions for $a_{i}$ and $\delta A_{i}$
in Eqs. (\ref{eq:En_01}, \ref{eq:E_n_01}), 
\begin{eqnarray}
a_{1} & = & a_{6}=\zeta^{-1}\tau_{\Pi}(e+P),\nonumber \\
a_{2} & = & \frac{1}{[1+C_{1}n_{1}^{2}+C_{2}(n_{2}^{2}+n_{3}^{2})]},\nonumber \\
a_{3} & = & \frac{1+C_{2}(n_{1}^{2}+n_{2}^{2}+n_{3}^{2})}{[1+C_{1}(n_{1}^{2}+n_{2}^{2})+C_{2}n_{3}^{2}][1+C_{1}n_{1}^{2}+C_{2}(n_{2}^{2}+n_{3}^{2})]},\nonumber \\
a_{4} & = & \frac{1+C_{2}(n_{1}^{2}+n_{2}^{2}+n_{3}^{2})}{[1+C_{1}(n_{1}^{2}+n_{2}^{2})+C_{2}n_{3}^{2}][1+C_{1}(n_{1}^{2}+n_{2}^{2}+n_{3}^{2})]},\nonumber \\
a_{5} & = & \frac{1+(C_{1}-c_{s}^{2})(n_{1}^{2}+n_{2}^{2}+n_{3}^{2})}{1+C_{1}(n_{1}^{2}+n_{2}^{2}+n_{3}^{2})},\nonumber \\
a_{7} & = & \frac{1}{C_{3}+C_{4}n_{1}^{2}+C_{5}(n_{2}^{2}+n_{3}^{2})},\nonumber \\
a_{8} & = & \frac{C_{3}+C_{5}(n_{1}^{2}+n_{2}^{2}+n_{3}^{2})}{[C_{3}+C_{4}(n_{1}^{2}+n_{2}^{2})+C_{5}n_{3}^{2}][C_{3}+C_{4}n_{1}^{2}+C_{5}(n_{2}^{2}+n_{3}^{2})]},\nonumber \\
a_{9} & = & \frac{C_{3}+C_{5}(n_{1}^{2}+n_{2}^{2}+n_{3}^{2})}{[C_{3}+C_{4}(n_{1}^{2}+n_{2}^{2})+C_{5}n_{3}^{2}][C_{3}+C_{4}(n_{1}^{2}+n_{2}^{2}+n_{3}^{2})]},\nonumber \\
a_{10} & = & \frac{\{C_{4}-c_{s}^{2}[(C_{3}-2)^{2}-(C_{3}-1)C_{4}]\}(n_{1}^{2}+n_{2}^{2}+n_{3}^{2})^{2}}{n_{0}^{2}[C_{3}+C_{4}(n_{1}^{2}+n_{2}^{2}+n_{3}^{2})]}\nonumber \\
 &  & +\frac{[C_{3}+C_{4}+(3C_{3}-4)c_{s}^{2}](n_{1}^{2}+n_{2}^{2}+n_{3}^{2})+C_{3}}{n_{0}^{2}[C_{3}+C_{4}(n_{1}^{2}+n_{2}^{2}+n_{3}^{2})]},
\end{eqnarray}
and
\begin{eqnarray}
\delta A_{1} & = & \delta A_{6}=\frac{n_{0}}{e+P}\delta\Pi-\frac{\zeta}{\tau_{\Pi}(e+P)}(n_{1}\delta u_{1}+n_{2}\delta u_{2}+n_{3}\delta u_{3}),\nonumber \\
\delta A_{2} & = & [1+C_{1}n_{1}^{2}+C_{2}(n_{2}^{2}+n_{3}^{2})]\delta u_{1}+(C_{1}-C_{2})n_{1}(n_{2}\delta u_{2}+n_{3}\delta u_{3})-\frac{c_{s}^{2}n_{0}n_{1}}{e+P}\delta e,\nonumber \\
\delta A_{3} & = & [1+C_{1}(n_{1}^{2}+n_{2}^{2})+C_{2}n_{3}^{2}]\delta u_{2}+(C_{1}-C_{2})n_{2}n_{3}\delta u_{3}-\frac{c_{s}^{2}n_{0}n_{2}}{e+P}\delta e,\nonumber \\
\delta A_{4} & = & [1+C_{1}(n_{1}^{2}+n_{2}^{2}+n_{3}^{2})]\delta u_{3}-\frac{c_{s}^{2}n_{0}n_{3}}{e+P}\delta e,\nonumber \\
\delta A_{5} & = & \delta A_{10}=\frac{c_{s}n_{0}}{e+P}\delta e,\nonumber \\
\delta A_{7} & = & [C_{3}+C_{4}n_{1}^{2}+C_{5}(n_{2}^{2}+n_{3}^{2})]\delta u_{1}+(C_{4}-C_{5})n_{1}(n_{2}\delta u_{2}+n_{3}\delta u_{3})\nonumber \\
 &  & +\frac{(C_{3}-2)c_{s}^{2}n_{1}n_{0}}{e+P}\delta e,\nonumber \\
\delta A_{8} & = & [C_{3}+C_{4}(n_{1}^{2}+n_{2}^{2})+C_{5}n_{3}^{2}]\delta u_{2}+(C_{4}-C_{5})n_{2}n_{3}\delta u_{3}+\frac{(C_{3}-2)c_{s}^{2}n_{2}n_{0}}{e+P}\delta e,\nonumber \\
\delta A_{9} & = & [C_{3}+C_{4}(n_{1}^{2}+n_{2}^{2}+n_{3}^{2})]\delta u_{3}+\frac{(C_{3}-2)c_{s}^{2}n_{3}n_{0}}{e+P}\delta e,
\end{eqnarray}
where we have defined
\begin{eqnarray}
C_{1} & = & 1-\frac{4\eta}{3\tau_{\pi}(e+P)}-\frac{\zeta}{\tau_{\Pi}(e+P)},\nonumber \\
C_{2} & = & 1-\frac{\eta}{\tau_{\pi}(e+P)},\nonumber \\
C_{3} & = & 1-\frac{\lambda}{\tau_{q}(e+P)},\nonumber \\
C_{4} & = & 1-\frac{\lambda}{\tau_{q}(e+P)}-\frac{4\eta}{3\tau_{\pi}(e+P)}-\frac{\zeta}{\tau_{\Pi}(e+P)},\nonumber \\
C_{5} & = & 1-\frac{\lambda}{\tau_{q}(e+P)}-\frac{\eta}{\tau_{\pi}(e+P)}-\frac{\gamma_{s}}{\tau_{\phi}(e+P)}.
\end{eqnarray}

\section{Expressions for $c_{2,3}$ in inequality (\ref{eq:LinearStable_4})
\label{sec:Expressionsbici}}

The expressions for $c_{2,3}$ in the inequality (\ref{eq:LinearStable_4})
are given by
\begin{eqnarray}
c_{1} & = & \sqrt{\frac{b_{1}^{1/2}\pm(b_{1}-b_{2})^{1/2}}{6(2\tau_{q}-\lambda^{\prime})\tau_{\pi}\tau_{\Pi}}},\textrm{or }-\sqrt{\frac{b_{1}^{1/2}\pm(b_{1}-b_{2})^{1/2}}{6(2\tau_{q}-\lambda^{\prime})\tau_{\pi}\tau_{\Pi}}},\nonumber \\
c_{2} & = & -3c_{1}^{4}[2\tau_{\pi}\tau_{\Pi}+(2\tau_{q}-\lambda^{\prime})(\tau_{\pi}+\tau_{\Pi})]+c_{1}^{2}\{6\gamma_{\|}\tau_{q}+(6\gamma_{\|}-8\gamma_{\perp})\tau_{\pi}\nonumber \\
 &  & +8\gamma_{\perp}\tau_{\Pi}+3c_{s}^{2}[2\tau_{\pi}\tau_{\Pi}+(3\lambda^{\prime}+2\tau_{q})(\tau_{\pi}+\tau_{\Pi})]\}-3c_{s}^{2}\gamma_{\|}\lambda^{\prime},\nonumber \\
c_{3} & = & -2c_{s}^{2}\lambda^{\prime}[(3\gamma_{\|}-4\gamma_{\perp})\tau_{\pi}+4\gamma_{\perp}\tau_{\Pi}]-18c_{1}^{4}(2\tau_{q}-\lambda^{\prime})\tau_{\pi}\tau_{\Pi}\nonumber \\
 &  & +4c_{1}^{2}[3c_{s}^{2}(3\lambda^{\prime}+2\tau_{q})\tau_{\pi}\tau_{\Pi}+2(3\gamma_{\|}-4\gamma_{\perp})\tau_{q}\tau_{\pi}+8\gamma_{\perp}\tau_{q}\tau_{\Pi}].
\end{eqnarray}
Note that here we have set $\chi_{e}^{\mu\nu}=0$, but the corresponding
formulas in Ref. \citep{Xie:2023gbo} contain nonzero $\chi_{e}^{\mu\nu}$.

\section{Derive inequalities (\ref{eq:LinearStable_3}, \ref{eq:LinearStable_4})
from (\ref{eq:LinearStable_1}, \ref{eq:LinearStable_2}) \label{sec:Stability}}

In this appendix, we will show that the inequalities (\ref{eq:LinearStable_3}),
(\ref{eq:LinearStable_4}) can be derived from (\ref{eq:LinearStable_1}),
(\ref{eq:LinearStable_2}). In the following calculations, we adopt
the notations (\ref{eq:LinearNotations0}), (\ref{eq:LinearNotations1}),
in which we have 
\begin{equation}
3\gamma_{\|}-4\gamma_{\perp}>0.\label{eq:BasicIneq2}
\end{equation}
 The inequalities (\ref{eq:LinearStable_2}), (\ref{eq:BasicIneq2})
will be frequently used.

For the inequality (\ref{eq:LinearStable_3}), we note that 
\begin{eqnarray}
b_{2} & = & 12c_{s}^{2}\lambda^{\prime}(2\tau_{q}-\lambda^{\prime})\tau_{\pi}\tau_{\Pi}[\tau_{\pi}(3\gamma_{\|}-4\gamma_{\perp})+4\gamma_{\perp}\tau_{\Pi}],\nonumber \\
b_{1}-b_{2} & = & 9(3\lambda^{\prime}+2\tau_{q})^{2}\tau_{\pi}^{2}\tau_{\Pi}^{2}c_{s}^{4}+4\tau_{q}^{2}[(3\gamma_{\|}-4\gamma_{\perp})\tau_{\pi}+4\gamma_{\perp}\tau_{\Pi}]^{2}\nonumber \\
 &  & +12(\lambda^{\prime2}+\lambda^{\prime}\tau_{q}+2\tau_{q}^{2})\tau_{\pi}\tau_{\Pi}[(3\gamma_{\|}-4\gamma_{\perp})\tau_{\pi}+4\gamma_{\perp}\tau_{\Pi}]c_{s}^{2}.
\end{eqnarray}
Using (\ref{eq:LinearStable_1}), (\ref{eq:LinearStable_2}), (\ref{eq:BasicIneq2}),
we find that $b_{2}>0$ and $b_{1}-b_{2}>0$, proving the inequality
(\ref{eq:LinearStable_3}).

To show the inequality (\ref{eq:LinearStable_4}), it is equivalent
to show $c_{2}c_{3}>0$. Straightforward calculation gives 
\begin{equation}
c_{2}c_{3}=f_{0}\pm f_{1}(b_{1}-b_{2})^{1/2},\label{eq:c2Tc3}
\end{equation}
where
\begin{eqnarray}
f_{0} & = & \frac{1}{9\tau_{\pi}^{3}\tau_{\Pi}^{3}(2\tau_{q}-\lambda^{\prime})^{3}}f_{0}^{(1)}f_{0}^{(2)},\nonumber \\
f_{1} & = & \frac{1}{9\tau_{\pi}^{3}\tau_{\Pi}^{3}(2\tau_{q}-\lambda^{\prime})^{3}}\left[f_{1}^{(0)}+c_{s}^{2}f_{1}^{(2)}+c_{s}^{4}f_{1}^{(4)}+c_{s}^{6}f_{1}^{(6)}\right],
\end{eqnarray}
with 
\begin{eqnarray}
f_{0}^{(1)} & = & 16\tau_{\pi}\gamma_{\perp}\tau_{\Pi}\left[3c_{s}^{2}\tau_{\Pi}(\lambda^{\prime2}+\tau_{q}\lambda^{\prime}+2\tau_{q}^{2})+2(3\gamma_{\|}-4\gamma_{\perp})\tau_{q}^{2}\right]\nonumber \\
 &  & +9c_{s}^{4}\tau_{\pi}^{2}\tau_{\Pi}^{2}(3\lambda^{\prime}+2\tau_{q})^{2}+12c_{s}^{2}\tau_{\pi}^{2}\tau_{\Pi}(3\gamma_{\|}-4\gamma_{\perp})(\lambda^{\prime2}+\tau_{q}\lambda^{\prime}+2\tau_{q}^{2})\nonumber \\
 &  & +4\tau_{\pi}^{2}\tau_{q}^{2}(3\gamma_{\|}-4\gamma_{\perp})^{2}+64\gamma_{\perp}^{2}\tau_{\Pi}^{2}\tau_{q}^{2},\nonumber \\
\nonumber \\
f_{0}^{(2)} & = & 72c_{s}^{4}\lambda^{\prime}\tau_{\pi}^{3}\tau_{\Pi}^{3}(3\lambda^{\prime}+2\tau_{q})+64\gamma_{\perp}^{2}\tau_{\Pi}^{3}\tau_{q}(\tau_{\pi}\lambda^{\prime}+\tau_{q}\lambda^{\prime}+2\tau_{q}^{2})\nonumber \\
 &  & +12c_{s}^{2}\tau_{\pi}\tau_{\Pi}^{3}\gamma_{\perp}(2\tau_{q}+\lambda^{\prime})(4\tau_{q}^{2}-\lambda^{\prime2}+8\lambda\tau_{\pi})\nonumber \\
 &  & +4(3\gamma_{\|}-4\gamma_{\perp})^{2}\tau_{\pi}^{3}\tau_{q}[\lambda^{\prime}\tau_{\Pi}+\tau_{q}(2\tau_{q}-\lambda^{\prime})]\nonumber \\
 &  & +(3\gamma_{\|}-4\gamma_{\perp})\tau_{\pi}\tau_{\Pi}\left\{ 3c_{s}^{2}\tau_{\pi}^{2}\tau_{\Pi}(2\tau_{q}+\lambda^{\prime})(4\tau_{q}^{2}-\lambda^{\prime2}+8\lambda^{\prime}\tau_{\Pi})\right.\nonumber \\
 &  & \quad\left.+16\gamma_{\perp}\tau_{\pi}\tau_{q}[2\lambda^{\prime}\tau_{\Pi}+\tau_{q}(2\tau_{q}-\lambda^{\prime})]+16\gamma_{\perp}\tau_{\Pi}\tau_{q}^{2}(2\tau_{q}-\lambda^{\prime})\right\} ,\nonumber \\
\nonumber \\
f_{1}^{(0)} & = & 8\tau_{q}^{2}[4\gamma_{\perp}(\tau_{\Pi}-\tau_{\pi})+3\tau_{\pi}\gamma_{\parallel}]^{2}\left\{ 4\gamma_{\perp}\tau_{\Pi}^{2}[\lambda^{\prime}\tau_{\pi}+\tau_{q}(2\tau_{q}-\lambda^{\prime})]\right.\nonumber \\
 &  & \quad\left.+\lambda^{\prime}\tau_{\pi}^{2}\tau_{\Pi}(3\gamma_{\parallel}-4\gamma_{\perp})+\tau_{\pi}^{2}\tau_{q}(3\gamma_{\parallel}-4\gamma_{\perp})(2\tau_{q}-\lambda^{\prime})\right\} ,\nonumber \\
\nonumber \\
f_{1}^{(2)} & = & 6\tau_{\pi}\tau_{\Pi}[\tau_{\pi}(3\gamma_{\parallel}-4\gamma_{\perp})+4\gamma_{\perp}\tau_{\Pi}]\nonumber \\
 &  & \times\left\{ \tau_{\pi}^{2}(3\gamma_{\parallel}-4\gamma_{\perp})[2\lambda^{\prime}\tau_{\Pi}(\lambda^{\prime2}+5\tau_{q}\lambda^{\prime}+10\tau_{q}^{2})+4\tau_{q}^{3}(\lambda^{\prime}+4\tau_{q})-3\lambda^{\prime3}\tau_{q}]\right.\nonumber \\
 &  & \quad\left.+4\gamma_{\perp}\tau_{\Pi}^{2}[2\lambda^{\prime}\tau_{\pi}(\lambda^{\prime2}+5\tau_{q}\lambda^{\prime}+10\tau_{q}^{2})+4\tau_{q}^{3}(\lambda^{\prime}+4\tau_{q})-3\lambda^{\prime3}\tau_{q}]\right\} ,\nonumber \\
\nonumber \\
f_{1}^{(4)} & = & 72[(3\gamma_{\parallel}-4\gamma_{\perp})\tau_{\pi}+4\tau_{\Pi}\gamma_{\perp}]\tau_{\pi}^{3}\tau_{\Pi}^{3}\lambda^{\prime}(5\lambda^{\prime2}+10\lambda^{\prime}\tau_{q}+8\tau_{q}^{2})\nonumber \\
 &  & +9[(3\gamma_{\parallel}-4\gamma_{\perp})\tau_{\pi}^{2}+4\tau_{\Pi}^{2}\gamma_{\perp}]\tau_{\pi}^{2}\tau_{\Pi}^{2}(2\tau_{q}-\lambda^{\prime})(2\tau_{q}+\lambda^{\prime})^{2}(3\lambda^{\prime}+2\tau_{q}),\nonumber \\
\nonumber \\
f_{1}^{(6)} & = & 216\lambda^{\prime}\tau_{\pi}^{4}\tau_{\Pi}^{4}(3\lambda^{\prime}+2\tau_{q})^{2}.
\end{eqnarray}
From the inequalities (\ref{eq:LinearStable_1}), (\ref{eq:LinearStable_2}),
(\ref{eq:BasicIneq2}), we have 
\begin{equation}
f_{0}>0.\label{eq:f0L0}
\end{equation}
Next we calculate 
\begin{eqnarray}
f_{0}^{2}-f_{1}^{2}(b_{1}-b_{2}) & = & (g_{0}+g_{2}c_{s}^{2}+g_{4}c_{s}^{2})G,
\end{eqnarray}
where 
\begin{eqnarray}
G & = & \frac{4\lambda^{\prime2}c_{s}^{4}}{9\tau_{\pi}^{3}\tau_{\Pi}^{3}(2\tau_{q}-\lambda^{\prime})^{3}}[\tau_{\pi}(3\gamma_{\parallel}-4\gamma_{\perp})+4\tau_{\Pi}\gamma_{\perp}]\nonumber \\
 &  & \times\left\{ \tau_{\Pi}^{2}\left[48\tau_{\pi}c_{s}^{2}\gamma_{\perp}\left(\lambda^{2}+\tau_{q}(\lambda+2\tau_{q})\right)+9\tau_{\pi}^{2}c_{s}^{4}(3\lambda+2\tau_{q})^{2}+64\gamma_{\perp}^{2}\tau_{q}^{2}\right]\right.\nonumber \\
 &  & \quad+4\tau_{\pi}\tau_{\Pi}(3\gamma_{\parallel}-4\gamma_{\perp})\left[3\tau_{\pi}c_{s}^{2}\left(\lambda^{2}+\tau_{q}(\lambda+2\tau_{q})\right)+8\gamma_{\perp}\tau_{q}^{2}\right]\nonumber \\
 &  & \quad\left.+4\tau_{\pi}^{2}\tau_{q}^{2}(3\gamma_{\parallel}-4\gamma_{\perp})^{2}\right\} ,\nonumber \\
\nonumber \\
g_{0} & = & 4\left\{ 4\gamma_{\perp}\tau_{\Pi}^{2}[\lambda^{\prime}\tau_{\pi}+\tau_{q}(2\tau_{q}-\lambda^{\prime})]+\tau_{\pi}^{2}[\tau_{q}(2\tau_{q}-\lambda^{\prime})+\lambda^{\prime}\tau_{\Pi}](3\gamma_{\|}-4\gamma_{\perp})\right\} \nonumber \\
 &  & \times[4\gamma_{\perp}\tau_{\Pi}+\tau_{\pi}(3\gamma_{\|}-4\gamma_{\perp})]^{2},\nonumber \\
\nonumber \\
g_{2} & = & 24\tau_{\pi}^{3}(3\gamma_{\|}-4\gamma_{\perp})\gamma_{\perp}\tau_{\Pi}^{2}[4\lambda^{\prime2}+16\lambda^{\prime}\tau_{\Pi}+(2\tau_{q}-\lambda^{\prime})^{2}]\nonumber \\
 &  & +96\tau_{\pi}\gamma_{\perp}^{2}\tau_{\Pi}^{4}[4\lambda^{\prime2}+(2\tau_{q}-\lambda^{\prime})^{2}]+48\gamma_{\perp}^{2}\tau_{\Pi}^{4}(2\tau_{q}+\lambda^{\prime})^{2}(2\tau_{q}-\lambda^{\prime})\nonumber \\
 &  & +768\lambda^{\prime}\gamma_{\perp}^{2}\tau_{\pi}^{2}\tau_{\Pi}^{4}+24\gamma_{\perp}\tau_{\pi}^{2}\tau_{\Pi}^{3}(3\gamma_{\|}-4\gamma_{\perp})[4\lambda^{\prime2}+(2\tau_{q}-\lambda^{\prime})^{2}]\nonumber \\
 &  & +24\gamma_{\perp}\tau_{\pi}^{2}\tau_{\Pi}^{2}(3\gamma_{\|}-4\gamma_{\perp})(2\tau_{q}+\lambda^{\prime})^{2}(2\tau_{q}-\lambda^{\prime})\nonumber \\
 &  & +6\tau_{\pi}^{4}\tau_{\Pi}(3\gamma_{\|}-4\gamma_{\perp})^{2}[4\lambda^{\prime2}+8\lambda^{\prime}\tau_{\Pi}+(2\tau_{q}-\lambda^{\prime})^{2}]\nonumber \\
 &  & +3\tau_{\pi}^{4}(3\gamma_{\|}-4\gamma_{\perp})^{2}(2\tau_{q}+\lambda^{\prime})^{2}(2\tau_{q}-\lambda^{\prime}),\nonumber \\
\nonumber \\
g_{4} & = & 72\lambda^{\prime}\tau_{\pi}^{2}\tau_{\Pi}^{2}\left\{ 4\gamma_{\perp}\tau_{\Pi}^{2}(3\lambda^{\prime}+2\tau_{q}+2\tau_{\pi})+\tau_{\pi}^{2}(3\gamma_{\|}-4\gamma_{\perp})[(3\lambda^{\prime}+2\tau_{q})+2\tau_{\Pi}]\right\} .
\end{eqnarray}
Again, we can find from the inequalities (\ref{eq:LinearStable_1}),
(\ref{eq:LinearStable_2}), (\ref{eq:BasicIneq2}) that 
\begin{equation}
G,g_{0},g_{2},g_{4}>0,
\end{equation}
which leads to 
\begin{equation}
f_{0}^{2}-f_{1}^{2}(b_{1}-b_{2})>0.\label{eq:f0mf1L0}
\end{equation}
Combing the results (\ref{eq:f0L0}) and (\ref{eq:f0mf1L0}), we obtain
\begin{equation}
c_{2}c_{3}=f_{0}\pm f_{1}(b_{1}-b_{2})^{1/2}>0,
\end{equation}
or the equivalent form, $c_{2}/c_{3}>0$, i.e. the inequality (\ref{eq:LinearStable_4}).

\bibliographystyle{h-physrev}
\bibliography{reference240502}

\begin{thebibliography}{100}

\bibitem{BRAHMS:2004adc}
BRAHMS, I.~Arsene {\em et~al.},
\newblock Nucl. Phys. A {\bf 757}, 1 (2005), nucl-ex/0410020.

\bibitem{PHENIX:2004vcz}
PHENIX, K.~Adcox {\em et~al.},
\newblock Nucl. Phys. A {\bf 757}, 184 (2005), nucl-ex/0410003.

\bibitem{STAR:2005gfr}
STAR, J.~Adams {\em et~al.},
\newblock Nucl. Phys. A {\bf 757}, 102 (2005), nucl-ex/0501009.

\bibitem{ALICE:2008ngc}
ALICE, K.~Aamodt {\em et~al.},
\newblock JINST {\bf 3}, S08002 (2008).

\bibitem{Israel:1979wp}
W.~Israel and J.~M. Stewart,
\newblock Annals Phys. {\bf 118}, 341 (1979).

\bibitem{Israel:1979}
W.~{Israel} and J.~M. {Stewart},
\newblock Proceedings of the Royal Society of London Series A {\bf 365}, 43
  (1979).

\bibitem{Baier:2007ix}
R.~Baier, P.~Romatschke, D.~T. Son, A.~O. Starinets, and M.~A. Stephanov,
\newblock JHEP {\bf 04}, 100 (2008), 0712.2451.

\bibitem{Denicol:2012cn}
G.~S. Denicol, H.~Niemi, E.~Molnar, and D.~H. Rischke,
\newblock Phys. Rev. D {\bf 85}, 114047 (2012), 1202.4551,
\newblock [Erratum: Phys.Rev.D 91, 039902 (2015)].

\bibitem{Bemfica:2017wps}
F.~S. Bemfica, M.~M. Disconzi, and J.~Noronha,
\newblock Phys. Rev. D {\bf 98}, 104064 (2018), 1708.06255.

\bibitem{Kovtun:2019hdm}
P.~Kovtun,
\newblock JHEP {\bf 10}, 034 (2019), 1907.08191.

\bibitem{Bemfica:2019knx}
F.~S. Bemfica {\em et~al.},
\newblock Phys. Rev. D {\bf 100}, 104020 (2019), 1907.12695,
\newblock [Erratum: Phys.Rev.D 105, 069902 (2022)].

\bibitem{Hoult:2020eho}
R.~E. Hoult and P.~Kovtun,
\newblock JHEP {\bf 06}, 067 (2020), 2004.04102.

\bibitem{Bemfica:2020zjp}
F.~S. Bemfica, M.~M. Disconzi, and J.~Noronha,
\newblock Phys. Rev. X {\bf 12}, 021044 (2022), 2009.11388.

\bibitem{Gavassino:2021kpi}
L.~Gavassino and M.~Antonelli,
\newblock Front. Astron. Space Sci. {\bf 8}, 686344 (2021), 2105.15184.

\bibitem{Rocha:2023ilf}
G.~S. Rocha, D.~Wagner, G.~S. Denicol, J.~Noronha, and D.~H. Rischke,
\newblock Entropy {\bf 26}, 189 (2024), 2311.15063.

\bibitem{Liang:2004ph}
Z.-T. Liang and X.-N. Wang,
\newblock Phys. Rev. Lett. {\bf 94}, 102301 (2005), nucl-th/0410079,
\newblock [Erratum: Phys.Rev.Lett. 96, 039901 (2006)].

\bibitem{Liang:2004xn}
Z.-T. Liang and X.-N. Wang,
\newblock Phys. Lett. B {\bf 629}, 20 (2005), nucl-th/0411101.

\bibitem{Gao:2007bc}
J.-H. Gao {\em et~al.},
\newblock Phys. Rev. C {\bf 77}, 044902 (2008), 0710.2943.

\bibitem{STAR:2017ckg}
STAR, L.~Adamczyk {\em et~al.},
\newblock Nature {\bf 548}, 62 (2017), 1701.06657.

\bibitem{STAR:2019erd}
STAR, J.~Adam {\em et~al.},
\newblock Phys. Rev. Lett. {\bf 123}, 132301 (2019), 1905.11917.

\bibitem{STAR:2022fan}
STAR, M.~S. Abdallah {\em et~al.},
\newblock Nature {\bf 614}, 244 (2023), 2204.02302.

\bibitem{Becattini:2007sr}
F.~Becattini, F.~Piccinini, and J.~Rizzo,
\newblock Phys. Rev. C {\bf 77}, 024906 (2008), 0711.1253.

\bibitem{Karpenko:2016jyx}
I.~Karpenko and F.~Becattini,
\newblock Eur. Phys. J. C {\bf 77}, 213 (2017), 1610.04717.

\bibitem{Xie:2017upb}
Y.~Xie, D.~Wang, and L.~P. Csernai,
\newblock Phys. Rev. C {\bf 95}, 031901 (2017), 1703.03770.

\bibitem{Li:2017slc}
H.~Li, L.-G. Pang, Q.~Wang, and X.-L. Xia,
\newblock Phys. Rev. C {\bf 96}, 054908 (2017), 1704.01507.

\bibitem{Sun:2017xhx}
Y.~Sun and C.~M. Ko,
\newblock Phys. Rev. C {\bf 96}, 024906 (2017), 1706.09467.

\bibitem{Shi:2017wpk}
S.~Shi, K.~Li, and J.~Liao,
\newblock Phys. Lett. B {\bf 788}, 409 (2019), 1712.00878.

\bibitem{Wei:2018zfb}
D.-X. Wei, W.-T. Deng, and X.-G. Huang,
\newblock Phys. Rev. C {\bf 99}, 014905 (2019), 1810.00151.

\bibitem{Xia:2018tes}
X.-L. Xia, H.~Li, Z.-B. Tang, and Q.~Wang,
\newblock Phys. Rev. C {\bf 98}, 024905 (2018), 1803.00867.

\bibitem{Vitiuk:2019rfv}
O.~Vitiuk, L.~V. Bravina, and E.~E. Zabrodin,
\newblock Phys. Lett. B {\bf 803}, 135298 (2020), 1910.06292.

\bibitem{Fu:2020oxj}
B.~Fu, K.~Xu, X.-G. Huang, and H.~Song,
\newblock Phys. Rev. C {\bf 103}, 024903 (2021), 2011.03740.

\bibitem{Ryu:2021lnx}
S.~Ryu, V.~Jupic, and C.~Shen,
\newblock Phys. Rev. C {\bf 104}, 054908 (2021), 2106.08125.

\bibitem{Lei:2021mvp}
A.~Lei, D.~Wang, D.-M. Zhou, B.-H. Sa, and L.~P. Csernai,
\newblock Phys. Rev. C {\bf 104}, 054903 (2021), 2110.13485.

\bibitem{Wu:2022mkr}
X.-Y. Wu, C.~Yi, G.-Y. Qin, and S.~Pu,
\newblock Phys. Rev. C {\bf 105}, 064909 (2022), 2204.02218.

\bibitem{Becattini:2013fla}
F.~Becattini, V.~Chandra, L.~Del~Zanna, and E.~Grossi,
\newblock Annals Phys. {\bf 338}, 32 (2013), 1303.3431.

\bibitem{Fang:2016vpj}
R.-h. Fang, L.-g. Pang, Q.~Wang, and X.-n. Wang,
\newblock Phys. Rev. C {\bf 94}, 024904 (2016), 1604.04036.

\bibitem{Liu:2021uhn}
S.~Y.~F. Liu and Y.~Yin,
\newblock JHEP {\bf 07}, 188 (2021), 2103.09200.

\bibitem{Liu:2021nyg}
Y.-C. Liu and X.-G. Huang,
\newblock Sci. China Phys. Mech. Astron. {\bf 65}, 272011 (2022), 2109.15301.

\bibitem{Fu:2021pok}
B.~Fu, S.~Y.~F. Liu, L.~Pang, H.~Song, and Y.~Yin,
\newblock Phys. Rev. Lett. {\bf 127}, 142301 (2021), 2103.10403.

\bibitem{Becattini:2021suc}
F.~Becattini, M.~Buzzegoli, and A.~Palermo,
\newblock Phys. Lett. B {\bf 820}, 136519 (2021), 2103.10917.

\bibitem{Yi:2021ryh}
C.~Yi, S.~Pu, and D.-L. Yang,
\newblock Phys. Rev. C {\bf 104}, 064901 (2021), 2106.00238.

\bibitem{Yi:2021unq}
C.~Yi, S.~Pu, J.-H. Gao, and D.-L. Yang,
\newblock Phys. Rev. C {\bf 105}, 044911 (2022), 2112.15531.

\bibitem{Yi:2023tgg}
C.~Yi {\em et~al.},
\newblock Phys. Rev. C {\bf 109}, L011901 (2024), 2304.08777.

\bibitem{Wu:2023tku}
X.-Y. Wu, C.~Yi, G.-Y. Qin, and S.~Pu,
\newblock EPJ Web Conf. {\bf 296}, 05008 (2024), 2312.09068.

\bibitem{Liu:2020dxg}
S.~Y.~F. Liu and Y.~Yin,
\newblock Phys. Rev. D {\bf 104}, 054043 (2021), 2006.12421.

\bibitem{Fu:2022myl}
B.~Fu, L.~Pang, H.~Song, and Y.~Yin,
\newblock (2022), 2201.12970.

\bibitem{Sun:2024isb}
J.-A. Sun and L.~Yan,
\newblock (2024), 2401.07458.

\bibitem{Fang:2023bbw}
S.~Fang, S.~Pu, and D.-L. Yang,
\newblock Phys. Rev. D {\bf 109}, 034034 (2024), 2311.15197.

\bibitem{ALICE:2021pzu}
ALICE, S.~Acharya {\em et~al.},
\newblock Phys. Rev. Lett. {\bf 128}, 172005 (2022), 2107.11183.

\bibitem{STAR:2023eck}
STAR,
\newblock (2023), 2303.09074.

\bibitem{Montenegro:2017rbu}
D.~Montenegro, L.~Tinti, and G.~Torrieri,
\newblock Phys. Rev. D {\bf 96}, 056012 (2017), 1701.08263,
\newblock [Addendum: Phys.Rev.D 96, 079901 (2017)].

\bibitem{Montenegro:2017lvf}
D.~Montenegro, L.~Tinti, and G.~Torrieri,
\newblock Phys. Rev. D {\bf 96}, 076016 (2017), 1703.03079.

\bibitem{Hattori:2019lfp}
K.~Hattori, M.~Hongo, X.-G. Huang, M.~Matsuo, and H.~Taya,
\newblock Phys. Lett. B {\bf 795}, 100 (2019), 1901.06615.

\bibitem{Fukushima:2020ucl}
K.~Fukushima and S.~Pu,
\newblock Phys. Lett. B {\bf 817}, 136346 (2021), 2010.01608.

\bibitem{Li:2020eon}
S.~Li, M.~A. Stephanov, and H.-U. Yee,
\newblock Phys. Rev. Lett. {\bf 127}, 082302 (2021), 2011.12318.

\bibitem{Gallegos:2021bzp}
A.~D. Gallegos, U.~G\"ursoy, and A.~Yarom,
\newblock SciPost Phys. {\bf 11}, 041 (2021), 2101.04759.

\bibitem{She:2021lhe}
D.~She, A.~Huang, D.~Hou, and J.~Liao,
\newblock (2021), 2105.04060.

\bibitem{Hongo:2021ona}
M.~Hongo, X.-G. Huang, M.~Kaminski, M.~Stephanov, and H.-U. Yee,
\newblock JHEP {\bf 11}, 150 (2021), 2107.14231.

\bibitem{Wang:2021ngp}
D.-L. Wang, S.~Fang, and S.~Pu,
\newblock Phys. Rev. D {\bf 104}, 114043 (2021), 2107.11726.

\bibitem{Wang:2021wqq}
D.-L. Wang, X.-Q. Xie, S.~Fang, and S.~Pu,
\newblock Phys. Rev. D {\bf 105}, 114050 (2022), 2112.15535.

\bibitem{Cao:2022aku}
Z.~Cao, K.~Hattori, M.~Hongo, X.-G. Huang, and H.~Taya,
\newblock PTEP {\bf 2022}, 071D01 (2022), 2205.08051.

\bibitem{Hu:2022azy}
J.~Hu,
\newblock (2022), 2209.10979.

\bibitem{Biswas:2023qsw}
R.~Biswas, A.~Daher, A.~Das, W.~Florkowski, and R.~Ryblewski,
\newblock (2023), 2304.01009.

\bibitem{Florkowski:2017ruc}
W.~Florkowski, B.~Friman, A.~Jaiswal, and E.~Speranza,
\newblock Phys. Rev. C {\bf 97}, 041901 (2018), 1705.00587.

\bibitem{Florkowski:2017dyn}
W.~Florkowski, B.~Friman, A.~Jaiswal, R.~Ryblewski, and E.~Speranza,
\newblock Phys. Rev. D {\bf 97}, 116017 (2018), 1712.07676.

\bibitem{Florkowski:2018myy}
W.~Florkowski, E.~Speranza, and F.~Becattini,
\newblock Acta Phys. Polon. B {\bf 49}, 1409 (2018), 1803.11098.

\bibitem{Weickgenannt:2019dks}
N.~Weickgenannt, X.-L. Sheng, E.~Speranza, Q.~Wang, and D.~H. Rischke,
\newblock Phys. Rev. D {\bf 100}, 056018 (2019), 1902.06513.

\bibitem{Bhadury:2020puc}
S.~Bhadury, W.~Florkowski, A.~Jaiswal, A.~Kumar, and R.~Ryblewski,
\newblock Phys. Lett. B {\bf 814}, 136096 (2021), 2002.03937.

\bibitem{Weickgenannt:2020aaf}
N.~Weickgenannt, E.~Speranza, X.-l. Sheng, Q.~Wang, and D.~H. Rischke,
\newblock Phys. Rev. Lett. {\bf 127}, 052301 (2021), 2005.01506.

\bibitem{Shi:2020htn}
S.~Shi, C.~Gale, and S.~Jeon,
\newblock Phys. Rev. C {\bf 103}, 044906 (2021), 2008.08618.

\bibitem{Speranza:2020ilk}
E.~Speranza and N.~Weickgenannt,
\newblock Eur. Phys. J. A {\bf 57}, 155 (2021), 2007.00138.

\bibitem{Bhadury:2020cop}
S.~Bhadury, W.~Florkowski, A.~Jaiswal, A.~Kumar, and R.~Ryblewski,
\newblock Phys. Rev. D {\bf 103}, 014030 (2021), 2008.10976.

\bibitem{Singh:2020rht}
R.~Singh, G.~Sophys, and R.~Ryblewski,
\newblock Phys. Rev. D {\bf 103}, 074024 (2021), 2011.14907.

\bibitem{Peng:2021ago}
H.-H. Peng, J.-J. Zhang, X.-L. Sheng, and Q.~Wang,
\newblock Chin. Phys. Lett. {\bf 38}, 116701 (2021), 2107.00448.

\bibitem{Sheng:2021kfc}
X.-L. Sheng, N.~Weickgenannt, E.~Speranza, D.~H. Rischke, and Q.~Wang,
\newblock Phys. Rev. D {\bf 104}, 016029 (2021), 2103.10636.

\bibitem{Hu:2021pwh}
J.~Hu,
\newblock Phys. Rev. D {\bf 105}, 076009 (2022), 2111.03571.

\bibitem{Weickgenannt:2022zxs}
N.~Weickgenannt, D.~Wagner, E.~Speranza, and D.~H. Rischke,
\newblock Phys. Rev. D {\bf 106}, 096014 (2022), 2203.04766.

\bibitem{Weickgenannt:2022jes}
N.~Weickgenannt, D.~Wagner, and E.~Speranza,
\newblock Phys. Rev. D {\bf 105}, 116026 (2022), 2204.01797.

\bibitem{Weickgenannt:2022qvh}
N.~Weickgenannt, D.~Wagner, E.~Speranza, and D.~H. Rischke,
\newblock Phys. Rev. D {\bf 106}, L091901 (2022), 2208.01955.

\bibitem{Wagner:2024fhf}
D.~Wagner, M.~Shokri, and D.~H. Rischke,
\newblock (2024), 2405.00533.

\bibitem{Gallegos:2020otk}
A.~D. Gallegos and U.~G\"ursoy,
\newblock JHEP {\bf 11}, 151 (2020), 2004.05148.

\bibitem{Garbiso:2020puw}
M.~Garbiso and M.~Kaminski,
\newblock JHEP {\bf 12}, 112 (2020), 2007.04345.

\bibitem{Becattini:2023ouz}
F.~Becattini, A.~Daher, and X.-L. Sheng,
\newblock Phys. Lett. B {\bf 850}, 138533 (2024), 2309.05789.

\bibitem{Hidaka:2022dmn}
Y.~Hidaka, S.~Pu, Q.~Wang, and D.-L. Yang,
\newblock Prog. Part. Nucl. Phys. {\bf 127}, 103989 (2022), 2201.07644.

\bibitem{Shi:2023sxh}
P.~Shi and H.~Xu-Guang,
\newblock Acta Phys. Sin. {\bf 72}, 071202 (2023).

\bibitem{Becattini:2024uha}
F.~Becattini {\em et~al.},
\newblock (2024), 2402.04540.

\bibitem{Hiscock:1985zz}
W.~A. Hiscock and L.~Lindblom,
\newblock Phys. Rev. D {\bf 31}, 725 (1985).

\bibitem{Hiscock:1987zz}
W.~A. Hiscock and L.~Lindblom,
\newblock Phys. Rev. D {\bf 35}, 3723 (1987).

\bibitem{Koide:2006ef}
T.~Koide, G.~S. Denicol, P.~Mota, and T.~Kodama,
\newblock Phys. Rev. C {\bf 75}, 034909 (2007), hep-ph/0609117.

\bibitem{Denicol:2008ha}
G.~S. Denicol, T.~Kodama, T.~Koide, and P.~Mota,
\newblock J. Phys. G {\bf 35}, 115102 (2008), 0807.3120.

\bibitem{Pu:2009fj}
S.~Pu, T.~Koide, and D.~H. Rischke,
\newblock Phys. Rev. D {\bf 81}, 114039 (2010), 0907.3906.

\bibitem{Brito:2020nou}
C.~V. Brito and G.~S. Denicol,
\newblock Phys. Rev. D {\bf 102}, 116009 (2020), 2007.16141.

\bibitem{Brito:2021iqr}
C.~V. Brito and G.~S. Denicol,
\newblock Phys. Rev. D {\bf 105}, 096026 (2022), 2107.10319.

\bibitem{Sarwar:2022yzs}
G.~Sarwar, M.~Hasanujjaman, J.~R. Bhatt, H.~Mishra, and J.-e. Alam,
\newblock (2022), 2209.08652.

\bibitem{Daher:2022wzf}
A.~Daher, A.~Das, and R.~Ryblewski,
\newblock Phys. Rev. D {\bf 107}, 054043 (2023), 2209.10460.

\bibitem{Xie:2023gbo}
X.-Q. Xie, D.-L. Wang, C.~Yang, and S.~Pu,
\newblock Phys. Rev. D {\bf 108}, 094031 (2023), 2306.13880.

\bibitem{Weickgenannt:2023btk}
N.~Weickgenannt,
\newblock Phys. Rev. D {\bf 108}, 076011 (2023), 2307.13561.

\bibitem{Shokri:2023rpp}
M.~Shokri and D.~H. Rischke,
\newblock Phys. Rev. D {\bf 108}, 096029 (2023), 2309.07003.

\bibitem{deBrito:2023vzv}
C.~V.~P. de~Brito, G.~S. Rocha, and G.~S. Denicol,
\newblock (2023), 2311.07272.

\bibitem{Fang:2024skm}
Z.~Fang, K.~Hattori, and J.~Hu,
\newblock (2024), 2402.18601.

\bibitem{Daher:2024bah}
A.~Daher, W.~Florkowski, R.~Ryblewski, and F.~Taghinavaz,
\newblock (2024), 2403.04711.

\bibitem{Krotscheck1978CausalityC}
E.~Krotscheck and W.~Kundt,
\newblock Communications in Mathematical Physics {\bf 60}, 171 (1978).

\bibitem{Heller:2022ejw}
M.~P. Heller, A.~Serantes, M.~Spali\'nski, and B.~Withers,
\newblock Phys. Rev. Lett. {\bf 130}, 261601 (2023), 2212.07434.

\bibitem{Gavassino:2023myj}
L.~Gavassino,
\newblock Phys. Lett. B {\bf 840}, 137854 (2023), 2301.06651.

\bibitem{Heller:2023jtd}
M.~P. Heller, A.~Serantes, M.~Spali\'nski, and B.~Withers,
\newblock (2023), 2305.07703.

\bibitem{Gavassino:2023mad}
L.~Gavassino, M.~M. Disconzi, and J.~Noronha,
\newblock Phys. Rev. Lett. {\bf 132}, 162301 (2024), 2307.05987.

\bibitem{Wang:2023csj}
D.-L. Wang and S.~Pu,
\newblock Phys. Rev. D {\bf 109}, L031504 (2024), 2309.11708.

\bibitem{Hoult:2023clg}
R.~E. Hoult and P.~Kovtun,
\newblock Phys. Rev. D {\bf 109}, 046018 (2024), 2309.11703.

\bibitem{Gavassino:2021owo}
L.~Gavassino,
\newblock Phys. Rev. X {\bf 12}, 041001 (2022), 2111.05254.

\bibitem{Hiscock:1983zz}
W.~A. Hiscock and L.~Lindblom,
\newblock Annals Phys. {\bf 151}, 466 (1983).

\bibitem{Olson:1990rzl}
T.~S. Olson,
\newblock Annals Phys. {\bf 199}, 18 (1990).

\bibitem{Gavassino:2021cli}
L.~Gavassino,
\newblock Class. Quant. Grav. {\bf 38}, 21LT02 (2021), 2104.09142.

\bibitem{Gavassino:2021kjm}
L.~Gavassino, M.~Antonelli, and B.~Haskell,
\newblock Phys. Rev. Lett. {\bf 128}, 010606 (2022), 2105.14621.

\bibitem{Landau:1980mil}
L.~D. Landau and E.~M. Lifshitz,
\newblock {\em {Statistical Physics, Part 1}}, Course of Theoretical Physics
  Vol.~5 (Butterworth-Heinemann, Oxford, 1980).

\bibitem{landau:1987Fluid}
L.~D. Landau and E.~M. Lifshitz,
\newblock {\em Fluid Mechanics}, Course of Theoretical Physics Vol.~6
  (Pergamon, New York, 1987).

\bibitem{Yagi:2005yb}
K.~Yagi, T.~Hatsuda, and Y.~Miake,
\newblock {\em {Quark-gluon plasma: From big bang to little bang}} (Cambridge
  University Press, Cambridge, 2005).

\bibitem{Almaalol:2022pjc}
D.~Almaalol, T.~Dore, and J.~Noronha-Hostler,
\newblock (2022), 2209.11210.

\bibitem{Pu:2011vr}
S.~Pu,
\newblock {\em {Relativistic fluid dynamics in heavy ion collisions}},
\newblock PhD thesis, Hefei, CUST, 2011, 1108.5828.

\bibitem{Gavassino:2023qnw}
L.~Gavassino and M.~Shokri,
\newblock Phys. Rev. D {\bf 108}, 096010 (2023), 2307.11615.

\bibitem{Becattini:2012tc}
F.~Becattini,
\newblock Phys. Rev. Lett. {\bf 108}, 244502 (2012), 1201.5278.

\bibitem{Becattini:2014yxa}
F.~Becattini, L.~Bucciantini, E.~Grossi, and L.~Tinti,
\newblock Eur. Phys. J. C {\bf 75}, 191 (2015), 1403.6265.

\bibitem{Becattini:2018duy}
F.~Becattini, W.~Florkowski, and E.~Speranza,
\newblock Phys. Lett. B {\bf 789}, 419 (2019), 1807.10994.

\bibitem{Gavassino:2024vyu}
L.~Gavassino, N.~Mullins, and M.~Hippert,
\newblock (2024), 2402.06776.

\bibitem{lasalle1961stability}
J.~LaSalle and S.~Lefschetz,
\newblock {\em Stability by Liapunov's Direct Method: With Applications},
  Mathematics in science and engineering Vol.~4 (Academic Press, New York,
  1961).

\bibitem{Gavassino:2023odx}
L.~Gavassino, M.~M. Disconzi, and J.~Noronha,
\newblock Phys. Rev. Lett. {\bf 132}, 222302 (2024), 2302.03478.

\end{thebibliography}

\end{document}